\documentclass[fleqn,usenatbib]{mnras}

\newcommand{\noopsort}[1]{}

\usepackage[T1]{fontenc}
\usepackage{ae,aecompl}
\usepackage{threeparttable}

\usepackage{graphicx}	
\usepackage{amsmath}	
\usepackage{amssymb}	
\usepackage{xcolor}

\title[total mass of UGC 8490 and UGC 9753]{General circular velocity relation of a test particle 
in a 3D gravitational potential: application to the rotation curves analysis and total mass 
determination of UGC 8490 and UGC 9753}

\author[P. Repetto et al.]{
P. Repetto,$^{1}$\thanks{E-mail: prsatch6@gmail.com}
Eric E. Mart\'{\i}nez-Garc\'{\i}a,$^{2}$
M. Rosado$^{3}$
and R. Gabbasov$^{4}$
\\
$^{1}$Via Gallaretta 42, Castelletto Orba, Alessandria, Italia, C.P. 15060\\
$^{2}$Cerrada del Rey 40A, Chimalcoyoc Tlalpan, M\'{e}xico D.F. C.P. 14630\\
$^{3}$Instituto de Astronom\'{\i}a, Universidad Nacional Autonoma de M\'{e}xico, Circuito de la Investigaci\'on Cient\'{\i}fica, \\
Ciudad Universitaria, M\'{e}xico, D.F., C.P. 04510\\
$^{4}$Instituto de Ciencias B\'asicas e Ingenier\'{\i}as, U.A.E.H., Carretera Pachuca-Tulancingo, M\'{e}xico, C.P. 42184}

\date{Accepted XXX. Received YYY; in original form ZZZ}

\pubyear{2015}

\begin{document}
\label{firstpage}
\pagerange{\pageref{firstpage}--\pageref{lastpage}}
\maketitle

\begin{abstract}
In this paper we derive a novel circular velocity relation for a test particle in a 3D gravitational potential applicable to every system of curvilinear coordinates, suitable to be reduced to orthogonal form. As an illustration of the potentiality of the determined circular velocity expression we perform the rotation curves analysis of UGC 8490 and UGC 9753 and we estimate the total and dark matter mass of these two galaxies under the assumption that their respective dark matter halos have spherical, prolate and oblate spheroidal mass distributions. We employ stellar population synthesis models and the total HI density map to obtain the stellar and HI+He+metals rotation curves of both galaxies. The subtraction of the stellar plus gas rotation curves from the observed rotation curves of UGC 8490 and UGC 9753 generates the dark matter circular velocity curves of both galaxies. We fit the dark matter rotation curves of UGC 8490 and UGC 9753 through the newly established circular velocity formula specialised to the spherical, prolate and oblate spheroidal mass distributions, considering the Navarro, Frenk and White, Burkert, Di Cintio, Einasto and Stadel dark matter halos. Our principal findings are the following: globally, cored dark matter profiles Burkert and Einasto prevail over cuspy Navarro, Frenk and White and Di Cintio. Also, spherical/oblate dark matter models fit better the dark matter rotation curves of both galaxies than prolate dark matter halos.
\end{abstract}




\section{Introduction}\label{sec:s1}

The possibility and validity to study the distribution of matter in galaxies, by means of the derived rotation curves (RCs), was established through the very important works of several authors, among which the more distinguished are \citet{Babcock1939, Wyse1942, Schmidt1957, Burbidge1959, Brandt1960}. The quoted papers approximated the mass distribution of some nearby \mbox{galaxies}, e.g., M31 and M33, through a series of oblate spheroids, or by means of thin disc geometries of homogeneous and non homogeneous density. The determined galaxy masses depend on the RCs spatial extent, the gaseous tracer used to build the analysed RCs, and on the specific geometrical assumptions considered. The results represent an adequate approximation to the actual total mass of the studied galaxies, as observed by the same authors. \citet{Brandt1962} extended the methodology of \citet{Brandt1960} (BD60) to obtain an estimation of a galaxy surface density from an hypothetical or observed RC, under the same geometrical assumptions of BD60. The derived surface density of M31 and NGC 5055, for instance, are concordant with previous determinations made by other authors, using different methods.

During the remaining part of the '60s and in the '70s, '80s and '90s, various authors perceived the necessity to trace the mass of galaxies through the extended neutral hydrogen RCs \citep{Roberts1975} to obtain optical RCs for larger radii and to increase the number of analysed RCs both in the optical and radio band \citep{Rubin1978, Rubin1980, Rubin1982, Rubin1985, Bosma1981}. The general trend of these periods was an attempt to collect as much as possible data on the RCs of galaxies improving the mass studies of the previous epochs in several aspects, such as for instance, the increase of the radial extension, the spectral and angular resolution of the measured circular velocities, the attempt to map the gravitational potential of the analysed galaxies with different gaseous and stellar tracers, the endeavour to provide more stringent clues about possible differences of the laws of rotation, mass contents and forms of the total gravitational potential among diverse galaxy morphologies. The paper of \citet{Mathewson1992} is one of the best examples of the new approach commented above, together with the influential papers of Rubin and collaborators that established, among other important properties of spiral galaxies RCs, the following points: the H$\alpha$ and HI RCs of spiral galaxies are flat at least up to $\sim$ 70 kpc, the flatness of the H$\alpha$ and HI RCs of spiral galaxies suggests that the luminous mass represents only a small fraction of the total mass, and the H$\alpha$ and HI RCs of the observed spiral galaxies show similar shapes for galaxies from Sa to Sc, indicating that the form of the entire gravitational potential is not dominated by the observed luminous mass. The more recent study of \citet{Sofue2013} shows, among other important results, that the RC of the Galaxy, derived by the author from the innermost to the outermost regions of the galactic disc, does not fall below 150 km s$^{-1}$ within the interval ranging from $\sim$ 0.1 up to 100 kpc. \citet{Sofue1997} was also probably the first to derive and use composite CO, HI and H$\alpha$ RCs to study the RCs behaviour and the kinematic of about 40 spiral galaxies in the local Universe.

Coeval and subsequent studies, aimed at determining the mass distribution in galaxies, focused principally on the possibility to separate the luminous matter component from the dark matter (DM), through the same RCs analysis or other associated techniques, to better determine the relative contribution of both components to the total RC. In the following we consider the two principal methods applied in the literature to divide the stellar and gas mass from the DM mass, and we recapitulate the more noticeable research works that apply these methods. The first procedure parameterizes the stellar component through a disc model with three parameters. The parameters are the stellar disc surface density and an horizontal and vertical scale length that can be obtained from the optical or infrared photometry of the studied galaxy applying standard methods and fitting formulas. The gas surface density can be obtained in a similar manner from the column density of the neutral hydrogen, and consequently the gaseous RC through an approximate relation between the surface density and the appropriate radial force. Several authors have successfully employed sophisticated variations to this first prescription to address the mass distribution issue in galaxies and other related questions such as for instance the discrepancy among the modeled cuspy DM inner density profile and the cored DM halo profile determined through the analysis of observed RCs, i.e., the cuspy/core problem \citep{Blais2001, Borriello2001, deBlok2002, Gentile2007, Spano2008, Swaters2012, Sofue2016}. 

The second strategy is based on a stellar population synthesis (SPS) analysis to derive the stellar disc mass of the studied galaxies. Although a detailed treatment of the SPS technique is not the intended purpose of this introduction, next we provide a concise description of the essential steps to attain the stellar surface densities from some observational and theoretical data related to the group of stars that reside in the analysed galaxy disc. In the basic picture, the stellar mass of a galaxy is determined through the product of the stellar mass-to-luminosity (M/L) ratio of the whole galaxy and the total stellar luminosity \citep{Walcher2011}. The stellar M/L according to \citet{Bell2001} shows a very sharp relation with optical and infrared colors, as a consequence, can be derived from the corresponding optical and infrared photometry. The total luminosity can be attained by fitting the Spectral Energy Distribution of the entire galaxy, that in turn is given by SPS models and their fundamental variable attributes, such as, for instance, the stellar population age, the star formation history, chemical composition, chemical elements abundance ratios and the initial mass function \citep{Courteau2014}. \citet{Zibetti2009} (ZB09), devised a new approach, based on optical and near infrared photometry, to obtain the stellar M/L at each pixel of the galaxy photometric images, to build an M/L 2D map and subsequently a 2D mass map using a vast Monte Carlo library of SPS models \citep{Bruzual2003} to establish the corresponding luminosity for every pixel. \citet{Martinez2017} (MG17) have recently ideated an alternative method to that of ZB09 to avoid the production of spurious filamentary mass portions in the resultant mass maps of ZB09, and also to mitigate the creation of other possible mass artefacts originated by the apparent superposition of dust lanes and spiral arms in the sample of analysed galaxies. Some authors, that determined the mass distribution of galaxies through the analysis of the observed RCs, employed the SPS strategy to derive the mass and circular velocity curve of the stellar discs considered, the results obtained are diverse and dependent on the specific assumptions of the particular approach they adopted, but in general concordant with the first prescription described above \citep{deBlok2008, Oh2008, Kuzio2008, Salucci2008, Denus2013} and also \citep{ Repetto2015, Repetto2017} (the latter paper RP17).           

In this article we examine the spherical, prolate and oblate spheroidal DM mass distributions, to ascertain which type of mass configuration is more adequate to reproduce the actual mass distribution of UGC 8490 and UGC 9753 as derived from the analysis of the observed RCs. We provide a new general formula capable to describe the circular motion of a particle in the gravitational fields generated by several distinct mass distributions, and in the present work we particularise that relation to the three specific geometries reported above. We detailed the derivation of the used circular velocity expression and other strictly related quantities in the corresponding addenda at the end of the manuscript and we introduce and apply the same formula throughout this work. As in RP17 we take advantage of the new approach of MG17 to build the stellar disc of UGC 8490 and UGC 9753, and we derive the total baryonic RCs in the same manner as in RP17. 

The rest of the paper is organised as follows: in section~(\ref{sec:s2}) we determine the type and number of stellar components that actually exist in the stellar disc of UGC 8490 and UGC 9753, in section~(\ref{sec:s3}) we outline the determination of the general circular velocity relation and every related quantity, in section~(\ref{sec:s6}) we detail the derivation of the baryonic RCs, as well as the HI and H$\alpha$ observed RCs, in section~(\ref{sec:s9}) we apply the devised circular velocity relation to the fit of the DM RCs of UGC 8490 and UGC 9753, considering a spherical, prolate and oblate DM distribution. In section~(\ref{sec:s10}) we present our results and compare with other studies. Sections~(\ref{sec:s12}) and~(\ref{sec:s14}) are dedicated to the discussion and the conclusions, respectively. In the appendixes we highlight the principal steps to derive the general circular velocity formula and each connected quantity.

\section{GALFIT 2D isophotal analysis of UGC 8490 and UGC 9753}\label{sec:s2}

The morphological type of UGC 8490 (NGC 5204) is SA(s)m according to NED, Hyperleda\footnote{http://leda.univ-lyon1.fr/} \citep{Makarov2014} and Simbad \citep{Wenger2000} astronomy database. The GALFIT \citep{Peng2002, Peng2010} 2D isophotal analysis detected two principal photometric components: a disc and a bulge. The fit of a 2D Sersic profile to the HST F606W image and to the Spitzer IRAC 3.6 $\micron$ image  determines the disc and bulge parameters tabulated in table~(\ref{tab:tb1}). UGC 9753 (NGC 5879) is a LINER/H II galaxy \citep{Filho2000}, and is classified as SAbc in NED, Hyperleda and Simbad database following the study of \citet{Ann2015}. The GALFIT 2D photometric study reveals three main features: a bright central spot, a disc and a bulge. We fit a 2D exponential disc profile to the SDSS {\it g} band image of UGC 9753 \citep{Abazajian2009}, after removing the luminous nuclear structure through a stellar point spread function extracted from the same SDSS image. We adjusted a 2D Sersic profile to the Spitzer IRAC 3.6 $\micron$ image to determine the parameters of the bulge component. The principal characteristics of the disc and bulge are listed in table~(\ref{tab:tb1}). It is worth to notice that we do not detect any bar component within the disc of UGC 8490, regardless of its morphological classification, and our results for both objects are concordant with the findings of the SG4 survey \citep{Buta2015}. The errors associated to each quantities listed in table~(\ref{tab:tb1}) are of the order of $\sim$1.2$\%$ and represent 1-sigma errors as estimated by GALFIT, during the fitting process, using Poisson statistic.       

\begin{table}
\centering
\caption{GALFIT 2D isophotal analysis results.}
\label{tab:tb1}
\begin{threeparttable}
\begin{tabular}{llllll}
\hline
UGC 8490 & m\tnote{(1)} & r$_d$\tnote{(2)}  & n\tnote{(3)} & q\tnote{(4)} & PA\tnote{(5)}\\
\hline
Disc & 7.6 & 0.9 & 0.93 & 0.65 & 2.63\\
Bulge & 10.4 & 0.6 & 0.37 & 0.65 & -48.42\\
\hline
UGC 9753 & m & r$_d$ & n & q & PA\\
\hline
Disc & 13.0 & 1.1 & -- & 0.35 & 56.47\\
Bulge & 5.86 & 0.7 & 1.46 & 0.36 & -9.36\\
\hline
\end{tabular}
\begin{tablenotes}
\item[1] Integrated magnitude.
\item[2] Disc scale length (kpc).
\item[3] Sersic index.
\item[4] Axial ratio.
\item[5] Position angle (degrees).
\end{tablenotes}
\end{threeparttable}
\end{table}
    
\section{Derivation of a general circular velocity equation}\label{sec:s3}

In this section we determine a circular velocity relation that can be specialised to every system of curvilinear coordinates able to be converted to orthogonal form. The full derivation of the new formula is given in appendix~(\ref{sec:Aa1}), in this section we provide the general equation and outline the principal assumptions. First of all it is important to explain that we do not present in this article a new method to determine the total mass of self-gravitating density distributions, instead we provide a novel circular velocity formula applicable to very different geometries and mass aggregates. The appendices from~(\ref{sec:Bb1}) to~(\ref{sec:Ee1}) exposed a particular procedure to apply the general circular velocity relation to the specific geometries considered in this work (e.g. prolate and oblate geometries). In a subsequent paper the first author of the present article (e.g. P.R.) will introduce a more comprehensive approach not confined to a particular transformation between cartesian and curvilinear coordinates, to a peculiar geometry of the density distribution and to a specific application. The following premises are fundamental to obtain the general circular velocity formula and constitute the essential basis of addendum~(\ref{sec:Aa1}). We consider a 3D density distribution without any specific geometrical connotation and whose total mass can generate its own gravitational potential well. We conceive a picture in which the total volume density is divided into 3D smaller portions, each one able to create its proper self-gravity. The presented sketch is adequate for every astrophysical system endowed with the suitable volumetric density distribution necessary to produce the corresponding gravitational potential and does not demand any particular restriction on the observed phenomenology and structural properties of the analysed object. The most direct consequence of the latter indication is that the derived circular velocity relation is applicable to every galactic and subgalactic system, regardless of the global structural morphology, internal substructures and the specific instant of dynamical evolution. An appropriate application of the provided velocity relation implies that the circular motions can be properly separated from the non circular motions by means of robust observational techniques. We consider the additional hypothesis that the observer position, at an specific time, can be described by the ordinary Euclidean metric tensor with the appropriate metric coefficients. The orthogonality condition implies that the metric tensor matrix is diagonal or can be converted to a diagonal matrix, and we assume that this condition is met for the mass systems considered. Under these assumptions, as detailed in appendix~(\ref{sec:Aa1}), the circular velocity relation reads:

\begin{equation}\label{eqn:ee1}
v^2_c(R) = \frac{4\pi G R}{\sqrt{g}g^{11}} \int_0^R \rho(r) \sqrt{g} dr
\end{equation} 

\noindent where $g^{11}$ is the first element of the inverse of the Euclidean metric tensor matrix, $g$ is the product of the squared diagonal metric tensor coefficients. The presented circular velocity expression can be applied to each gravitating system in which the circular motions dominate or can be separated from the dispersive motions, such as for instance to study the rotation law and total mass of galactic discs, DM halos, gas clouds associations, circumnuclear galactic discs, protoplanetary discs and gas clouds rotating around compact objects, to illustrate some examples. Other possible applications of equation~(\ref{eqn:ee1}) are the following: the creation of initial conditions of self-gravitating masses of very different geometries supported by rotation in numerical simulations of isolated and interacting galaxy discs and to describe the dynamical evolution of various types of subgalactic systems, the derivation of potential, density and surface density triplets of rotating and self-gravitating density distributions, the analysis of 2D velocity fields to quantify the non circular motions, with the possibility to use very diverse geometrical configurations for the rotation law, the creation of synthetic 2D velocity fields to analyse artificial RCs originated by distinct mass distributions with the possibility to add a phenomenological motivated non circular velocity component as a result of a comparison with observations. 

The principal advantage of this relation is the possibility to deal with different geometries defined by transformations among Cartesian and curvilinear coordinates. A possible drawback is the requirement of orthogonality that is not always straightforward to attain, depending on the specific curvilinear coordinates transformation. The tensorial form of the circular velocity expression is dictated by the necessity to emcompass a broader range of geometries and correspondingly a larger number of possible astrophysical phenomena that involve the kinematics of collisional and collisionless fluids. The specific problem addressed in this work is only an example of the possible applications of equation~(\ref{eqn:ee1}), however, a larger number of issues can be potentially studied. In the next section we apply equation~(\ref{eqn:ee1}) to two particular geometries adequate to address the observational problem treated in this paper.

\subsection{Specific geometries under study: prolate and oblate spheroidal mass distributions}\label{sec:s4}

In this study we are principally interested in the determination of the total gravitating mass of galaxies, therefore we have to specialise the circular velocity relation defined by equation~(\ref{eqn:ee1}) to some particular geometry, that could represent a satisfactory approximation to the total galactic mass distribution or to the mass distribution of only a constituent of a galaxy, such as for instance the DM halo. The methodology we have employed in our previous work RP17, to estimate the total mass of galaxies derives the stellar and gas masses from SPS models and the application of photometric and spectroscopic techniques, respectively, consequently the sole unknown quantity is represented by the DM halo mass and can be determined by the DM RC fit, as showed in RP17. In appendix~(\ref{sec:Bb1}) we detail the necessary steps to obtain the prolate and oblate circular velocities from equation~(\ref{eqn:ee1}), whereas in this section we report the resultant expressions. The rotation velocity of a DM particle under the influence of a prolate DM mass distribution is determined by the following relation:

\begin{align}\label{eqn:ee2}      
v^2_c(R_p) &= \frac{4\pi G}{R_p f(\eta)} \int_0^{R_p} \rho(r_p) r^2_p dr_p
\end{align}

\noindent where $r_p=\frac{r}{a}$ is the 1D prolate radius and $a$ is the minor axis of the considered prolate spheroid. The quantity $f(\eta)=\frac{\cos^2{\eta}}{b^2}+\frac{\sin^2{\eta}}{a^2}$ is the prolate aspherical factor, $b$ is the major axis of the prolate spheroid and $\eta$ is connected to the inclination of the galaxy along the line of sight. A similar expression holds for the rotation velocity of a DM particle in a DM oblate mass distribution, the exact relation is:

\begin{align}\label{eqn:ee3}      
v^2_c(R_o) &= \frac{4\pi G}{R_o h(\eta)} \int_0^{R_o} \rho(r_o) r^2_o dr_o
\end{align}

\noindent where $r_o=\frac{r}{b}$ is the 1D oblate radius and $b$ is the major axis of the oblate spheroid. The quantity $h(\eta)=\frac{\cos^2{\eta}}{a^2}+\frac{\sin^2{\eta}}{b^2}$ is the oblate aspherical factor, $a$ and $\eta$ have the same meaning of the corresponding symbols in the prolate geometry. The supposition of the prevalence of circular motions in the context of the DM halo kinematics seems reasonable for a dynamically relaxed system in virial equilibrium, furthermore, according to the standard theoretical view \citep{Blumenthal1986} (BM86), DM is a pressureless fluid and it seems to be a dominant gravitating component in some galaxies. These largely known properties of DM should prevent the formation of significant non circular motions associated to the DM component.
 
\subsubsection{Total masses of prolate and oblate spheroids}\label{sec:s5}

Expressions to estimate the total mass distribution of prolate and oblate spheroids are determined computing the volume elements of both geometries through their respective Jacobians. The resultant volume elements of prolate and oblate mass distributions are provided by the relations $dV_p=r^2 a^2 b \sin{\eta}\, dr d\psi d\eta$ and \mbox{$dV_o=r^2 b^2 a \sin{\eta}\,dr d\psi d\eta$}, respectively. The corresponding total masses are given by the following prescriptions:

\begin{align}\label{eqn:ee4}
M(R_p) &= 4\pi a^2 b \int_0^{R_p} \rho(r_p) r^2_p dr_p    &   [M_T]_p &= \lim_{R_p\to\infty} M(R_p)\nonumber\\
\vspace{0.5cm}
M(R_o) &= 4\pi b^2 a \int_0^{R_o} \rho(r_o) r^2_o dr_o    &   [M_T]_o &= \lim_{R_o\to\infty} M(R_o)
\end{align}

\noindent where as stated above $a$ and $b$ represent the minor and major axes of the prolate and oblate spheroids and the total masses are computed in the ideal limit of an infinite radial extent. The DM masses of UGC 8490 and UGC 9753 are determined from the DM RCs fit and the total masses represent the DM masses plus the baryonic masses. The stellar and gas masses are determined according to the information reported in section~(\ref{sec:s7}). The connection between the prolate and oblate circular velocity relations and the corresponding mass relations are expressed by $v^2_c(R_p)=\frac{M(R_p)G}{a^2 b R_p f(\eta)}$ and $v^2_c(R_o)=\frac{M(R_o)G}{a b^2 R_o h(\eta)}$. These relations resemble the corresponding spherical recipes that relate squared circular velocity and mass, and additionally reveal that the DM RCs fit is indeed a DM mass fit. The latter observation is actually a largely accepted hypothesis for spherical mass distributions, and it supports partially the usage of the DM RCs fit as a tool to obtain information about the total mass of the galactic system, considering also the prolate and oblate spheroidal mass distributions.        

\section{Observational test}\label{sec:s6}

We apply the prolate and oblate circular velocity expressions defined in section~(\ref{sec:s4}) to determine the total mass of UGC 8490 and UGC 9753. The total baryonic mass is obtained through methods that are independent of the DM RCs fit, as well as of the analysis related to the total H$\alpha$ and HI RCs, therefore the prolate and oblate rotation velocity relations were employed to perform the fit of the DM RCs of UGC 8490 and UGC 9753 to obtain the respective DM masses. The total mass is then determined by the sum of the DM mass and the baryonic mass. We explore three type of mass distributions for the DM halo of both galaxies: spherical, prolate and oblate. In the next sections we illustrate the derivation of the stellar and HI+He+metals RCs and the methodology utilised to establish the total stellar and gas masses, then we proceed to the DM RCs fit and to contrast our findings with the results of other authors.

\subsection{Stellar and HI+He+metals RCs}\label{sec:s7}

\begin{figure*}
\centering
\includegraphics[width=1.0\hsize]{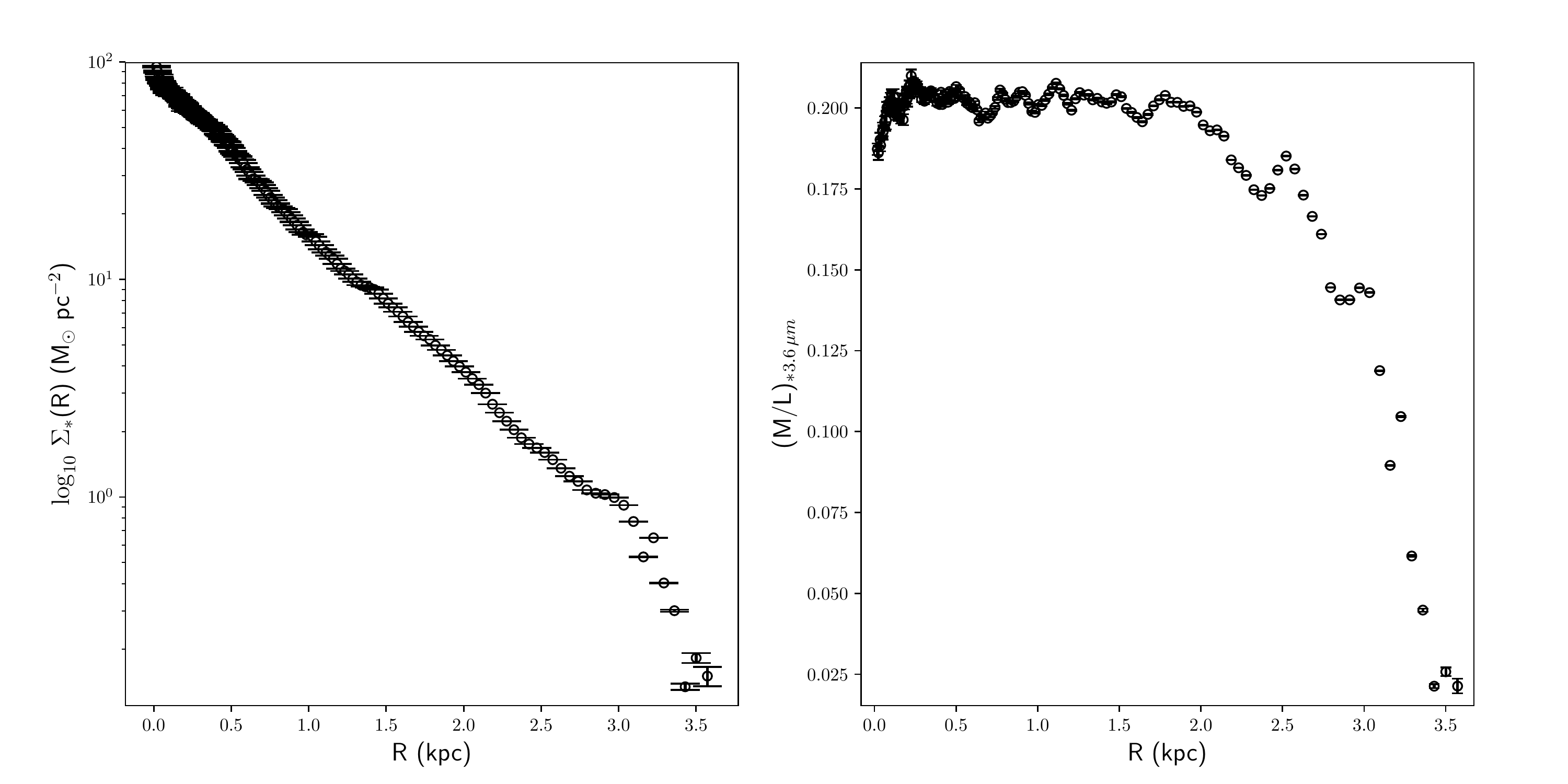}
\caption[f1.pdf]{{\it{Left}}: Stellar surface density of UGC 8490. We employ the MG17 method to build a 2D mass map of this galaxy and to obtain the stellar surface density and then the azimuthal average, using a position angle of 171.4$\degr$ and an inclination of 50.3$\degr$ as deprojection parameters. {\it{Right}}: M/L radial profile of UGC 8490 in the 3.6$\micron$ band.}
~\label{fig1}
\end{figure*}

\begin{figure*}
\centering
\includegraphics[width=1.0\hsize]{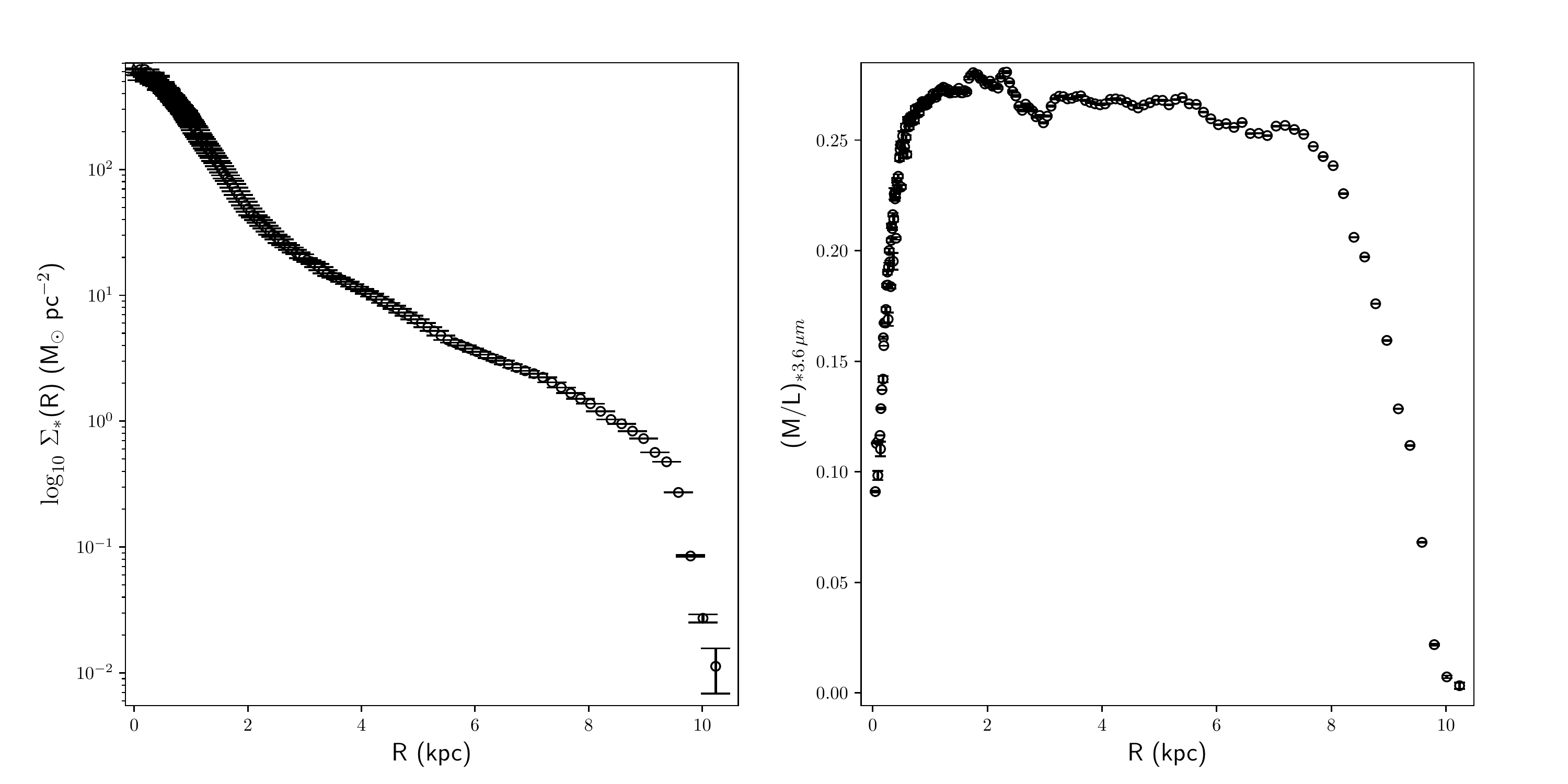}
\caption[f2.pdf]{Same information for UGC 9753 for both panels. The deprojection parameters used are a position angle of 2.3$\degr$ and an inclination of 70$\degr$.}
~\label{fig2}
\end{figure*}

\begin{figure*}
\centering
\includegraphics[width=1.0\hsize]{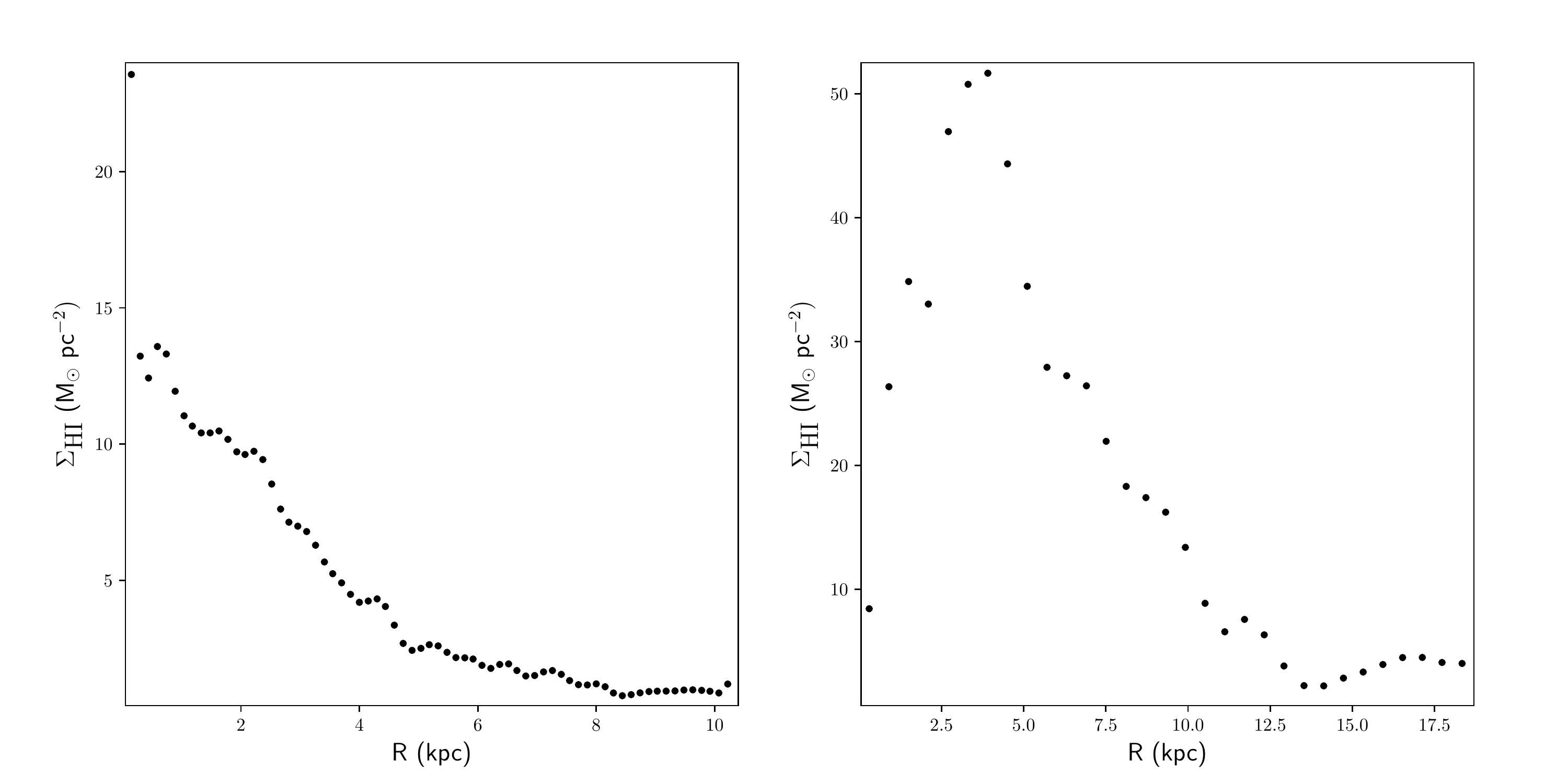}
\caption[f3.pdf]{{\it{Left}}: HI surface density of UGC 8490. As described in section~(\ref{sec:s7}) we use the ELLINT task of the GIPSY package to derive the HI surface density. The deprojection parameters used to obtain the HI surface density from the HI total density map of UGC 8490 through the ELLINT task, are a position angle of 171.4$\degr$ and an inclination of 50.3$\degr$ as tabulated in table~(\ref{tab:tb2}). {\it{Right}}: Same information for UGC 9753, with the exception that the position angle and inclination used for deprojection are 2.3$\degr$ and 70$\degr$, respectively.}
~\label{fig3}
\end{figure*}

We use SPS models to obtain the stellar surface densities and M/L radial curves of UGC 8490 and UGC 9753 showed in figure~(\ref{fig1}) and~(\ref{fig2}). We employ the MG17 Bayesian method in a similar way as described in RP17. We use the optical-NIR branch of the MAGPHYS library \citep{Cunha2008}, adopting the~\citet{Bruzual2003} (BC03) SPS models. For the photometry we make use of the SDSS DR12~\citep{Alam2015}, $g$ and $i$-band mosaics and the 3.6$\micron$ IRAC image from the S$^4$G~\citep{Sheth2010}. The maxima of 
the M/L curves of UGC 8490 and UGC 9753 displayed in figure~(\ref{fig1}) and~(\ref{fig2}) are $\sim$0.21 and $\sim$0.28, respectively, and 
are concordant with the corresponding values reported in table 4 of \citet{deBlok2008} (BK08) (e.g. M/L$\sim$0.23-0.31) for the sample of dwarf irregular galaxies studied by these authors, specifically NGC 2366, IC 2574 and DDO 154. For the same three galaxies the M/L values tabulated in table 3 of BK08 and table 3 and 4 of \citet{Oh2008} (OH08) (e.g. M/L$\sim$0.32-0.44) are a factor among $\sim$0.11-0.16 larger with respect to the maxima M/L values of UGC 8490 and UGC 9753. The difference is likely due to the diverse initial mass function used in our work when compared to the quoted studies. Despite of the dissimilarity among the M/L values, described above, the stellar masses of UGC 8490 and UGC 9753 are consistent with the corresponding stellar masses of NGC 2366, IC 2574 and DDO 154 listed in table 3 of BK08 (e.g. 2.6$\times$10$^8$, 1.1$\times$10$^9$, 2.6$\times$10$^7$ M$_{\odot}$, respectively, to compare with the values of stellar masses of UGC 8490 and UGC 9753 computed in section~(\ref{sec:s9}), separately, 2.1$\times$10$^8$ and 3.6$\times$10$^9$ M$_{\odot}$, also considering the dissimilar photometric diameters in the K$_s$ band). The conclusion is that the M/L discrepancy between our work and the study of BK08 and OH08 does not affects the results of the present work. The remaining galaxies studied by BK08 are barred spirals with a grand design spiral structure and the M/L values listed in table 3 and 4 of BK08 are consistent with the M/L values obtained by MG17 for corresponding large spiral galaxies, in the near infrared, particularly with M/L in the range $\sim$0.3-0.9, as it is evident both from figure 17 and table 1 of MG17. The ROTMOD \citep{Casertano1983} (CA83) task of the Groningen Image Processing System (GIPSY) package \citep{vanderHulst1992} was used to determine the stellar RCs of both galaxies. The ROTMOD routine computes a RC from an input surface density data set, using the recipe proposed by CA83. The error bars associated to the stellar RCs of UGC 8490 and UGC 9753 correspond to $\pm$17$\%$ of the stellar RC of UGC 8490 and to $\pm$16$\%$ of the stellar RC of UGC 9753. The HI surface densities were derived from the HI total density maps publicly available together with the HI velocity fields and data cubes by the Westerbork observations of neutral Hydrogen in Irregular and Spiral galaxies (WHISP) survey \citep{vanderHulst2001}. The GIPSY task ELLINT, subdivides the HI total density map in concentric elliptical rings and integrates the HI intensity over each elliptical subdivision to obtain the corresponding HI surface density, considering, as underlying supposition, a linear relation between the HI flux and mass. The HI surface density curves of UGC 8490 and UGC 9753 are displayed in figure~(\ref{fig3}). The HI+He+metals RCs are the HI RCs multiplied by the factor 1.33 that takes into account the primordial helium and metals fraction. The stellar and HI+He+metals RCs represent the total known amount of baryonic mass of UGC 8490 and UGC 9753, because the molecular hydrogen was not detected in the interstellar medium of UGC 8490 \citep{Leroy2005}, and we are not aware of any research work that reveals the presence of molecular hydrogen in the interstellar medium of UGC 9753. The HI flux errors of UGC 8490 are estimated by \citet{Jozsa2007} of the order of $\sim$ 10$\%$, whereas \citet{Springob2005} determined the HI flux errors of UGC 9753 to be $\sim$ 5$\%$. The corresponding errors of the HI rotational velocities are for both galaxies $\sim$ 5 km s$^{-1}$.

\subsection{H$\alpha$ and HI RCs}\label{sec:s8}

\begin{figure*}
\centering
\includegraphics[width=1.0\hsize]{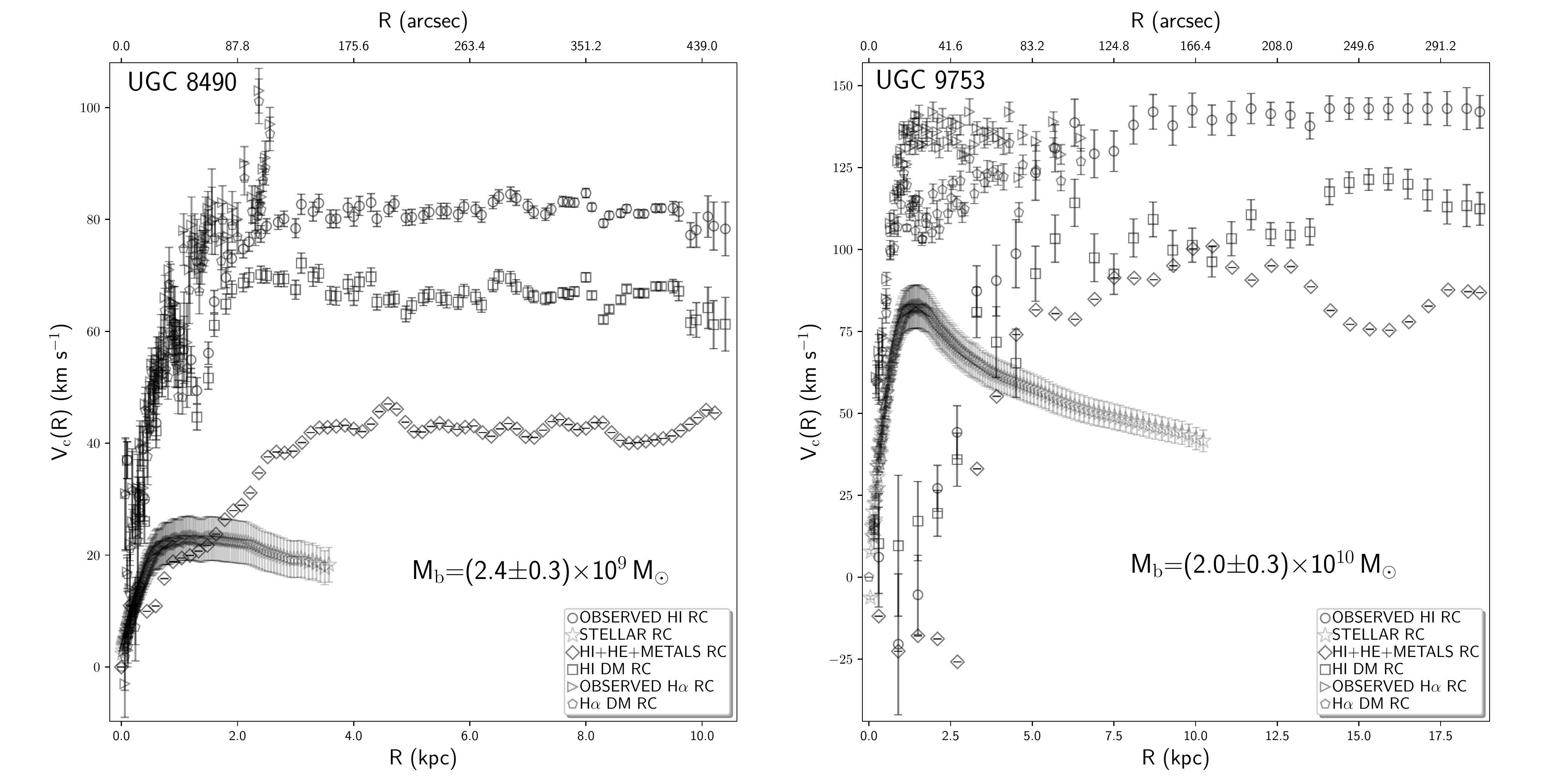}
\caption[f4.pdf]{{\it{Left}}: Observed HI RC (empty circles), HI DM RC (empty squares), HI+He+metals RC (empty diamonds), stellar RC (empty stars), observed H$\alpha$ RC (empty right triangles) and H$\alpha$ DM RC (empty pentagon) of UGC 8490. {\it{Right}}: Same information for UGC 9753. The derivation of the observed RCs, the DM RCs, the gas and stellar RCs is explained in sections~\ref{sec:s8} and~\ref{sec:s7}.}
~\label{fig4}
\end{figure*}

The H$\alpha$ RCs of UGC 8490 and UGC 9753 are available at the Vizier database and form part of the Ghassendi H$\alpha$ Survey of Spirals (GHASP) \citep{Garrido2002, Amram2002, Epinat2008}. The derivation of the H$\alpha$ RCs, the data reduction, and other important details are provided in the cited papers. In the next paragraphs we describe the determination of the HI RC from the HI velocity fields of both galaxies. We employ the GIPSY task ROTCUR \citep{Begeman1989} to obtain the total HI RCs from the HI velocity fields. As indicated by the kinematic and photometric study of \citet{Sicotte1997} the HI disc of UGC 8490 presents a severe distorsion beyond $\sim$3 kpc from the center of rotation. The same authors perform a tilted rings analysis of the HI disc of UGC 8490 to determine the HI RC and the radial variation of position angle and inclination. \citet{Swaters2009} noted the position angle and inclination variation of 65$\degr$ and 20$\degr$, separately, during the determination of the HI RC. As far as we know, there are no additional peculiarities of the HI disc of UGC 8490 capable to affect the derivation of its HI RC. On the other hand, UGC 9753 does not seem to present any relevant characteristic in its HI disc that could interfere with the determination of the HI RC basic parameters. We perform the tilted rings study \citep{Rogstad1971} of the HI disc of UGC 8490 through the ROTCUR routine beginning with a set of initial parameters close to the photometric parameters tabulated in table~(\ref{tab:tb1}), in particular the position angle and the inclination. The first guess of the systemic velocity was adopted from the NED database and it is concordant with the systemic velocity listed in the Hyperleda and Simbad databases. The procedure used to determine the RCs parameters of UGC 8490 and UGC 9753 is the following: at each iteration we let vary only one parameter and we fix all the remaining parameters to their respective initial guesses, we allow to vary in order the kinematic center, the systemic velocity, the position angle, the inclination, and the circular velocity. The variation of one parameter creates a vector of values of the same parameter, the average of that vector represents the optimal parameter value to use in the next step where we let vary a different parameter.
Several iterations could be necessary, for all or some parameter, to obtain a RC clear of systematic errors, that could arise due to the inaccurate determination of one or more parameters. The partial twist of the HI disc of UGC 8490 could primarily affect the determination of the position angle and the inclination that should display a more noticeable random behaviour. The actual radial variation of the position angle and inclination of UGC 8490 is moderately irregular, and their averaged values still allow to build a reasonable law of rotation. The resultant RC seems to be not so much affected by the presence of a strong warp in its HI disc as we can see from figure~(\ref{fig4}) and as we commented above the radial trend of the position angle and inclination of UGC 8490 are not so anomalous, as a consequence we do not display the resultant radial curves. The best parameters values for UGC 8490 are reported in table~(\ref{tab:tb2}). The same methodology was applied to derive the HI RC of UGC 9753 (see table~(\ref{tab:tb2})). The HI RC of UGC 9753 and UGC 8490, as well as the corresponding baryonic RCs, are displayed in figure~(\ref{fig4}). The HI velocity field analysis clearly shows the predominance of circular motions, because the velocity peaks of the residual velocity fields, obtained through the subtraction of the synthetic velocity maps from the observed velocity fields of both galaxies, do not present radial velocities higher than 20 km s$^{-1}$. In table~(\ref{tab:tb2}) we also compare the HI RC parameters of UGC 8490 with those of \citet{Sicotte1997} and \citet{Swaters2003}. We are not aware of any kinematic study of UGC 9753, as a consequence we can not compare the kinematic parameters obtained for this galaxy with any other previous determination. In figure~(\ref{fig4}) the error bars of the observed HI RCs of both galaxies are determined through the ROTCUR task, whereas the DM H$\alpha$ and HI RCs as well as the respective error bars are originated by the relations $v_{DM_{H\alpha}}=\left[v^2_{H\alpha}-v^2_{\star}\right]^{\frac{1}{2}}$ and $v_{DM_{HI}}=\left[v^2_{HI}-(v^2_{\star}+v^2_{HI+He+metals})\right]^{\frac{1}{2}}$, separately. The methodology employed to construct the DM H$\alpha$ and HI RCs is in part based on the original suggestions of \citet{Carignan1985} and \citet{Kravtsov1998}.
             
\begin{table}
\centering
\caption{HI RCs of UGC 8490 and UGC 9753: optimal parameters}
\label{tab:tb2}
\begin{threeparttable}
\begin{tabular}{p{1.2cm}p{1.3cm}p{1.3cm}p{1.3cm}p{1.3cm}}
\hline
Parameters & U8490\tnote{(5)} & U8490\tnote{(6)} & U8490\tnote{(7)} & U9753\tnote{(5)}\\
\hline
$\alpha$\tnote{(1)} & 13 29 27.0 & 13 29 52.5  & 13 29 36.4 & 15 09 42.6\\

$\delta$\tnote{(1)} & 58 24 44.0 & 58 23 32.6 & 58 25 12 & 56 59 52.0\\

V$_{sys}$\tnote{(2)} & 207.6$\pm$3.0  & 200.0$\pm$4.0  & 208.0\tnote{(8)} & 771.0$\pm$5.0\\

PA\tnote{(3)} & 171.4$\pm$3.0 & 203.0\tnote{(8)} & 170.0\tnote{(8)} & 2.3$\pm$1.7\\

INCL\tnote{(4)} & 50.3$\pm$2.0 & 47.3\tnote{(8)} & 50.0\tnote{(8)}& 70.0$\pm$2.5\\
\hline
\end{tabular}
\begin{tablenotes}
\item[1] Kinematic centre (h.m.s., d.m.s.).
\item[2] Systemic velocity (km s$^{-1}$).
\item[3] Kinematic position angle (degrees).
\item[4] Kinematic inclination (degrees).
\item[5] This work.
\item[6] \citet{Sicotte1997}.
\item[7] \citet{Swaters2003}.
\item[8] No error bars provided by the authors.
\end{tablenotes}
\end{threeparttable}
\end{table}

\section{DM and total mass for spherical, prolate and oblate geometries}\label{sec:s9}

We use the spherical, prolate and oblate circular velocity relations provided in section~(\ref{sec:s4}) to fit the DM H$\alpha$ and HI RCs of UGC 8490 and UGC 9753, and subsequently obtain the DM and total mass distribution of both galaxies. The total mass is the sum of the DM mass plus the stellar and gas mass. The total stellar mass is determined from the stellar surface densities, through the expression $M_{\star}(R)=2\pi\int_0^R \Sigma_{\star}(r)rdr$, where the integral extends up to the utmost radial range of the stellar surface densities displayed in figure~(\ref{fig1}) and~(\ref{fig2}). The principal supposition to compute the total stellar mass is that the stars move prevalently in circular orbits. We obtain a total stellar mass of (2.1$\pm$0.4)$\times$10$^8$ M$_{\odot}$ inside a radius of 3.6 kpc for UGC 8490, and a total stellar mass of (3.6$\pm$0.6)$\times$10$^9$ M$_{\odot}$ within a radius of 10.2 kpc for UGC 9753. The total gas mass is estimated employing the baryonic Tully-Fisher \citep{Tully1977} relation as provided by \citet{McGaugh2012} (MG12). According to this author the total baryonic mass is connected to the stellar and gas mass by the expression $M_b=M_{\star}+M_{gas}$. In the specific case of UGC 8490 and UGC 9753, the total gas mass of both galaxies coincide with the HI+He+metals gas mass, because both objects seem to be devoided of molecular hydrogen (see section~\ref{sec:s7} for references regarding the latter statement). We apply the baryonic Tully-Fisher relation of MG12 considering a circular velocity maximum of 85 km s$^{-1}$, and 143 km s$^{-1}$, for UGC 8490 and UGC 9753, separately. The total baryonic mass is (2.4$\pm$0.3)$\times$10$^9$ M$_{\odot}$ and (2.0$\pm$0.3)$\times$10$^{10}$ M$_{\odot}$ for UGC 8490 and UGC 9753, respectively. The total gas mass is given by the expression $M_{gas}=M_b-M_{\star}$, and the neutral hydrogen mass is provided by the relation $M_{HI}=\frac{M_{gas}}{1.33}$. We obtain a total gas mass of (2.2$\pm$0.3)$\times$10$^9$ M$_{\odot}$ and a neutral hydrogen mass of (1.7$\pm$0.2)$\times$10$^9$ M$_{\odot}$ for UGC 8490, a total gas mass of (1.6$\pm$0.2)$\times$10$^{10}$ M$_{\odot}$ and a neutral hydrogen mass of (1.2$\pm$0.1)$\times$10$^{10}$ M$_{\odot}$ for UGC 9753.

Generally the DM RC fits present only two free parameters, the DM halo mass and scale radius, therefore the axes of symmetry of the prolate and oblate spheroids, that enter in the corresponding prolate and oblate circular velocity relations, are fixed quantities whose constant values are defined in terms of some constraints related, for instance, to the condition of energy conservation. The latter seems a reasonable assumption, because according to the $\Lambda$CDM paradigm, the DM can be considered as a non dissipative fluid (BM86) and as a consequence its total energy is a conserved quantity. In appendix~(\ref{sec:Ee1}) we highlight the main assumptions and steps necessary to determine the expressions of the major and minor axes of symmetry of the prolate and oblate spheroids considered in this study. 
We delineate the degree of prolateness and oblateness of the assumed DM halos of UGC 8490 and UGC 9753 through the parameter $T=1-\left(\frac{a}{b}\right)^2$, where $a$ is the minor axis and $b$ is the major axis of symmetry, adapting the definition of triaxiality provided by \citet{Franx1991} to the prolate and oblate spheroidal biaxial geometry (setting $c=0$). A spheroidal prolate mass distribution has $T >$ 0.67, whereas a spheroidal oblate mass distribution presents $T <$ 0.3. In particular from our analysis, we have found that for a prolate spheroidal mass distribution $T=$0.75 and $T=$0.88 for UGC 8490 and UGC 9753, separately, and for an oblate mass distribution $T=$ 0.21 and $T=$0.28 for UGC 8490 and UGC 9753, respectively. We estimated the average uncertainties of the major and minor axes of the prolate and oblate spheroids employing equation~(\ref{eqn:ep1}) and the mean errors in the DM masses and radii as determined through the fits to the DM RCs of both galaxies. We found that for the prolate spheroidal mass distribution the average uncertainties are $\sim$ 0.02 and $\sim$ 0.05 for the minor and major axes, respectively, and for the oblate spheroidal geometry the mean errors are $\sim$ 0.01 and $\sim$ 0.04 for the minor and major axes, separately. The expressions that define the range of variation of the major and minor axes of the prolate and oblate spheroidal mass distributions derived in appendix~(\ref{sec:Ee1}) and the definition of $T$ can be justified in the standard cosmological framework, because they are functions of the principal parameters that determine galaxies evolution, such as mass, radius and the Hubble parameter at the present time. The formulation presented in appendix~(\ref{sec:Ee1}) is supported by cosmological theoretical works, e.g., \citet{Suto2016}.

The DM RCs fitting procedure is similar to the one applied in \citet{Repetto2015} and RP17, the principal difference in this work is represented by the introduction of the prolate and oblate circular velocity fit. We employ the {\bf minuit} fitting routine through the ROOT package \citep{ReneBrun} to adjust the spherical, prolate and oblate circular velocity relations to the DM RCs of UGC 8490 and UGC 9753. We use five DM halos density laws, specifically, the Navarro, Frenk and White (NFW) DM density profile \citep{Navarro1996}, the Burkert (BKT) DM density law \citep{Burkert1995}, the Di Cintio (DCN) DM density profile \citep{DiCintio2014}, the Einasto (EIN) DM density law \citep{Einasto1965} and the Stadel (STD) DM density profile \citep{Stadel2009}. As we have already mentioned above, the DM halo is characterized by two parameters, the mass and the scale radius that define its most important properties, therefore the third parameter of the EIN and STD DM halos is fixed to a specific arbitrary value for each fitting step of the DM halo mass and scale radius, and this procedure is repeated until the two free parameters converge to an acceptable solution. The DM halos mass and scale radius vary in the intervals $[10^8, 10^{13}]$ M$_{\odot}$ and $[0.1, 100.0]$ kpc, respectively. The best fit EIN and STD shape parameters for the spherical, prolate and oblate mass distributions in general take values within the range $[2.3, 4.0[$, therefore the corresponding solutions present a cored behaviour. The DCN DM halo shows a prevalent cuspy trend with inner slope greater than 0.35 for the spherical, prolate and oblate spheroidal mass distributions fitted to the DM H$\alpha$ and HI RCs of UGC 8490 and UGC 9753. 
The solutions determined through the spherical, prolate and oblate spheroidal fits to the DM RCs, were judged by means of three principal tools: the reduced $\chi^2_r$, the $\chi^2$ sigma contours and the estimated distance to the assumed global minimum (EDM). The $\chi^2_r$ is the $\chi^2$ normalized for the number of degrees of freedom (NDF), where the $\chi^2$ is defined in the usual manner as the square of the observed data minus the model weighted by the respective errors. The NDF of the H$\alpha$ DM RC fits of UGC 8490 and UGC 9753 are 74 and 57, separately, whereas the NDF of the HI DM RC fits are 68 and 30, respectively.  The $\chi^2$ sigma contours are $\chi^2$ error levels generated by the different values of the variable parameters that arise during the minimization process of the same $\chi^2$. The EDM measures the separation among the solution and the resulting global minimum, consequently an EDM close to zero should denote a successful $\chi^2$ minimization. During the fitting process, the {\bf minuit} routine builds a grid of local minima finding a new minimum after a certain number of iterations and grid steps, through a Monte Carlo search procedure. At the same time, a Metropolis approach is used to seek the global minimum, measuring \mbox{iteratively} the probability to find the deepest minimum among all the local minima originated by the minimization process. The probability is expressed as the inverse exponential of the minimum deepness, as a consequence, the lowest possible probability indicates a likely global minimum. The various EDMs, originated in the course of the minimization process, coincide wih the measured probabilities. In general the criteria that must be met simultaneously to obtain ideal fitting results are a $\chi^2_r$ near unity, an EDM close to zero and a resultant solution within the innermost error ellipse in the $\chi^2$ sigma contours plane. The results obtained for UGC 8490 and UGC 9753 describe with adequate accuracy the behaviour of the DM halo mass and scale radius at the present time, however as it is evident from the presented solutions, the three requirements mentioned above are never met at the same time.

\section{Results}\label{sec:s10}

\begin{figure*}
\centering
\includegraphics[width=1.0\hsize]{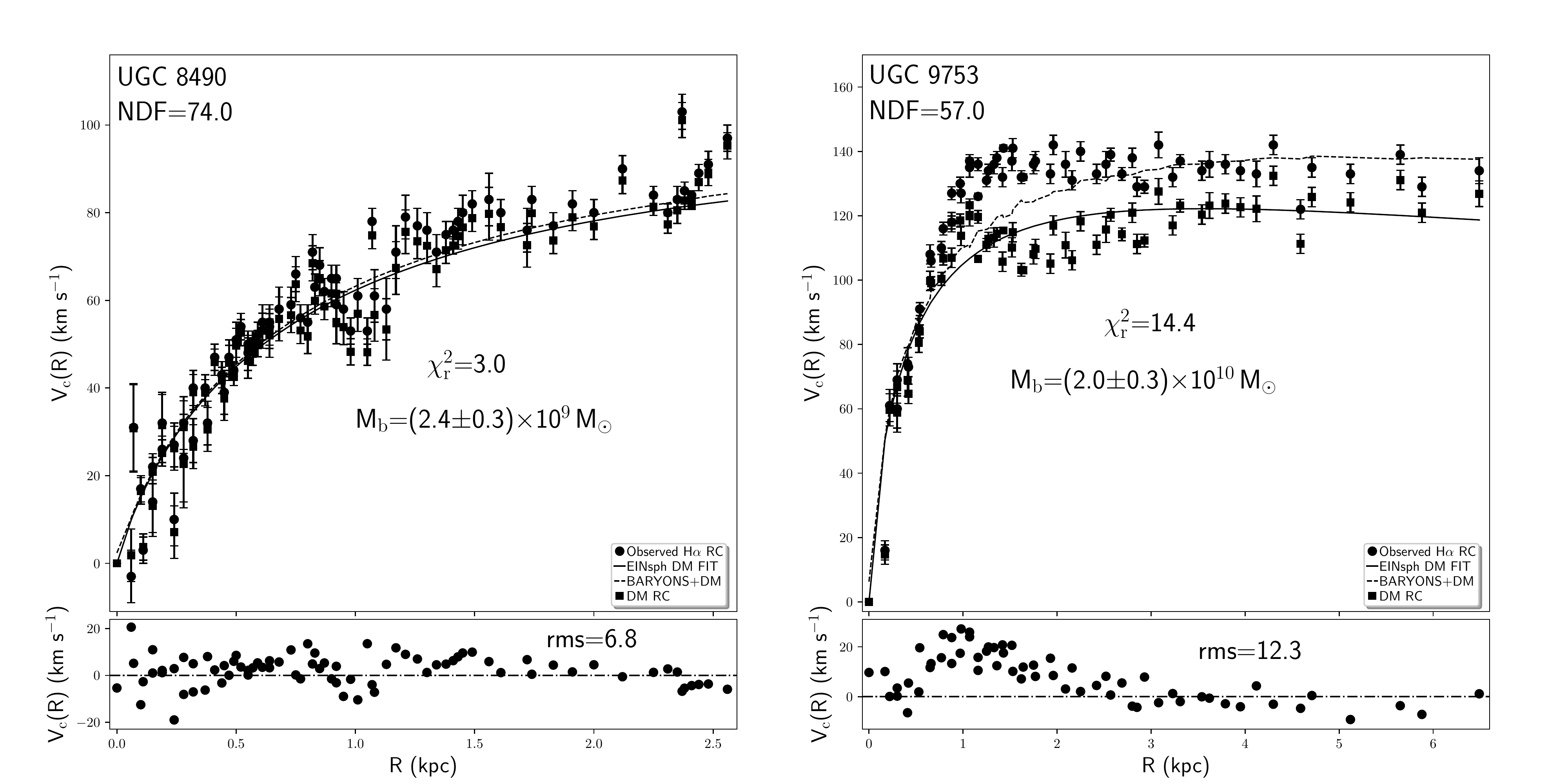}
\caption[f5.pdf]{{\it{Left}}: EIN spherical fit (solid line) to the H$\alpha$ DM RC of UGC 8490. The observed H$\alpha$ RC is displayed together with the H$\alpha$ DM RC of the same galaxy. The dashed line denotes the EIN DM fit to the H$\alpha$ DM RC plus the stellar RC. The residuals in the lower panel represent the subtraction of the squared EIN DM fit to the H$\alpha$ DM RC plus the squared stellar RC, from the squared observed H$\alpha$ RC. {\it{Right}}: Same information for UGC 9753.}
~\label{fig5}
\end{figure*}

\begin{figure*}
\centering
\includegraphics[width=1.0\hsize]{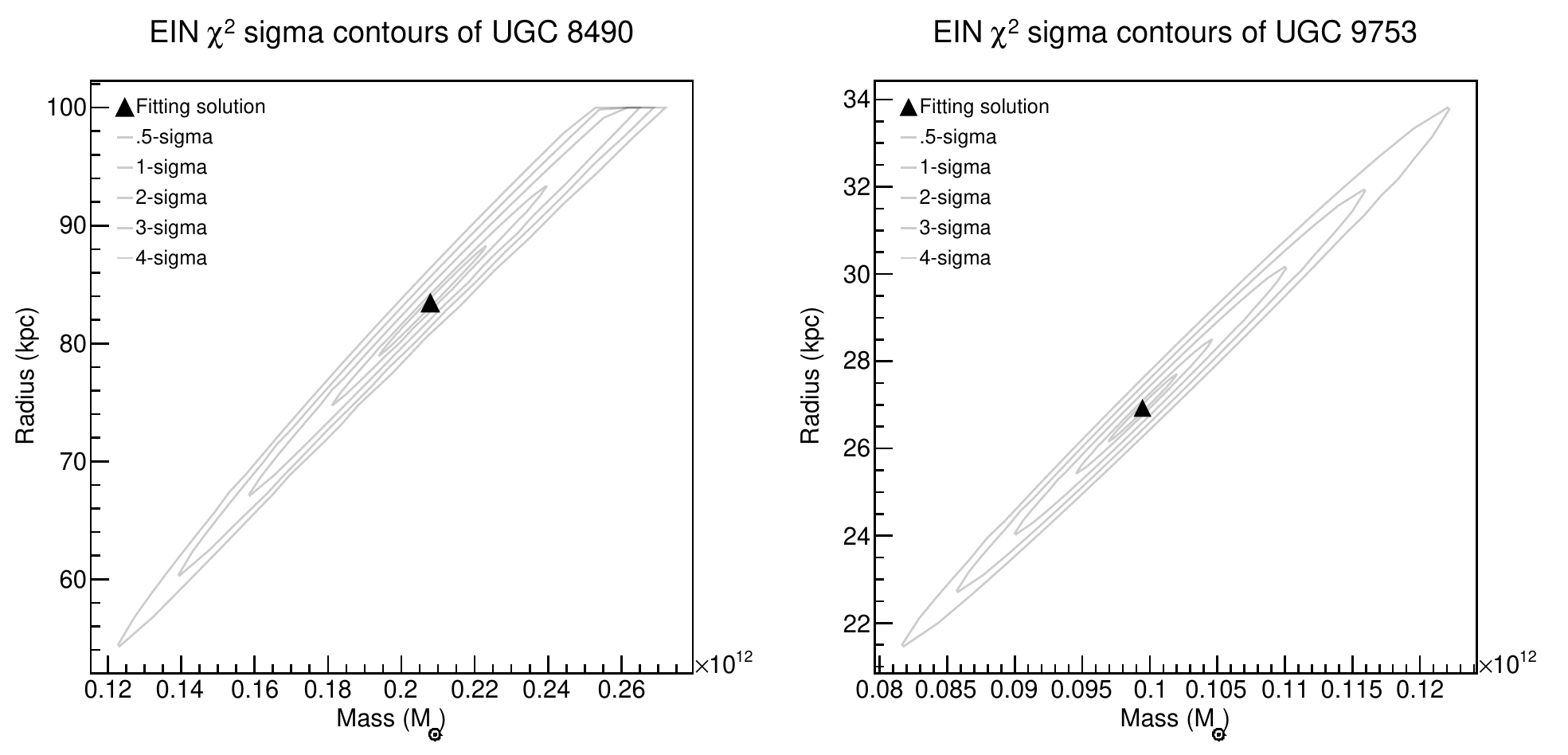}
\caption[f6.pdf]{{\it{Left}}: EIN $\chi^2$ sigma contours of the EIN spherical DM fit to the H$\alpha$ DM RC of UGC 8490. The contours ellipses represent five $\chi^2$ error contour levels. The solution is indicated by a black filled triangle and lies near or within the 1-sigma contour level. The optimal solution is expected to stay inside the 0.5-sigma contour level. {\it{Right}}: Same information for UGC 9753.}
~\label{fig6} 
\end{figure*}

\begin{figure*}
\centering
\includegraphics[width=1.0\hsize]{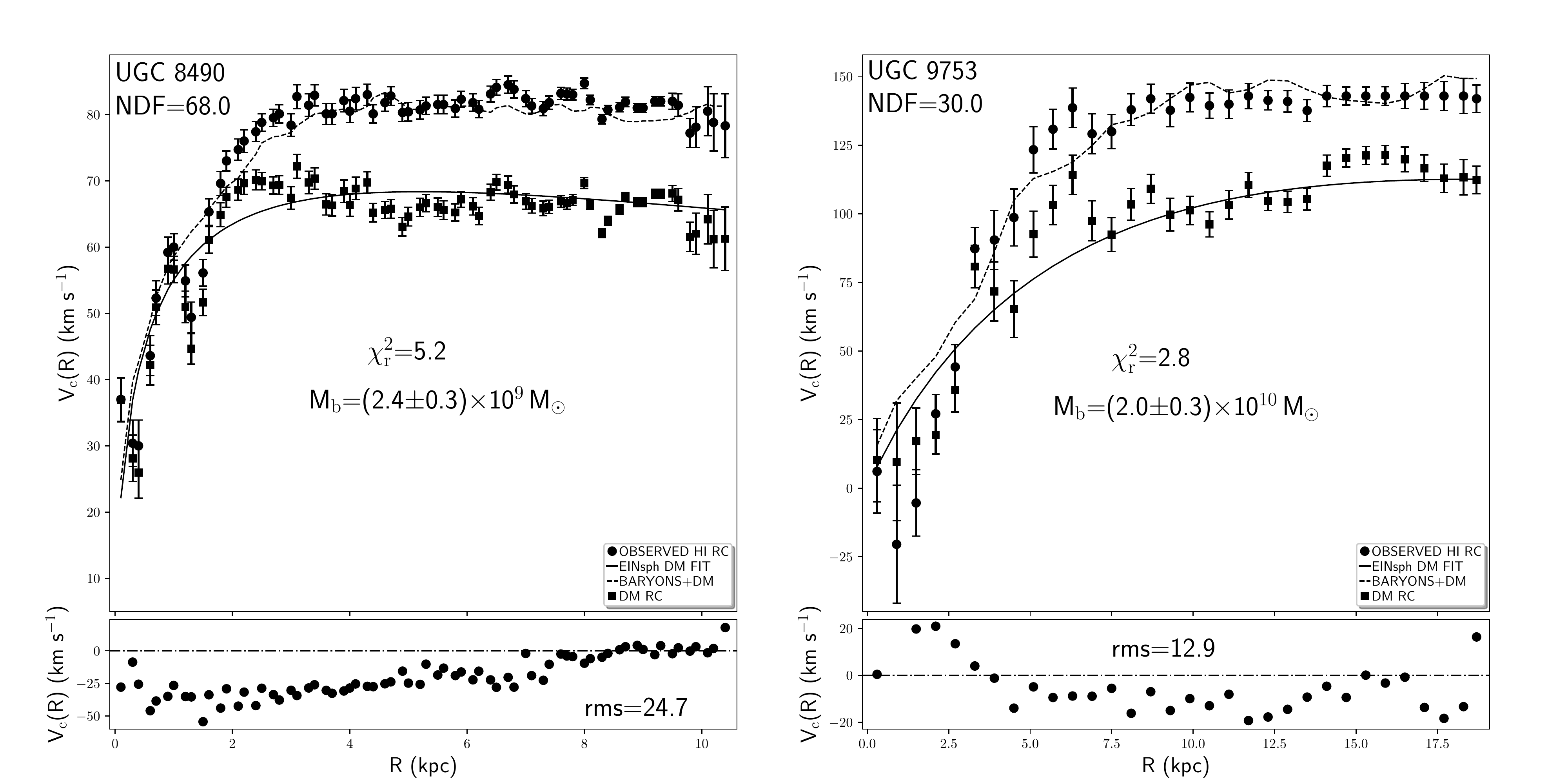}
\caption[f7.pdf]{{\it{Left}}: EIN spherical fit (solid line) to the HI DM RC of UGC 8490. The observed HI RC is displayed together with the HI DM RC of the same galaxy. The dashed line denotes the EIN DM fit to the HI DM RC plus the stellar RC and the HI+He+metals RC. The residuals in the lower panel represent the subtraction of the squared EIN DM fit to the HI DM RC plus the squared stellar RC and the squared HI+He+metals RC, from the squared observed HI RC. {\it{Right}}: Same information for UGC 9753.}
~\label{fig7}
\end{figure*}

\begin{figure*}
\centering
\includegraphics[width=1.0\hsize]{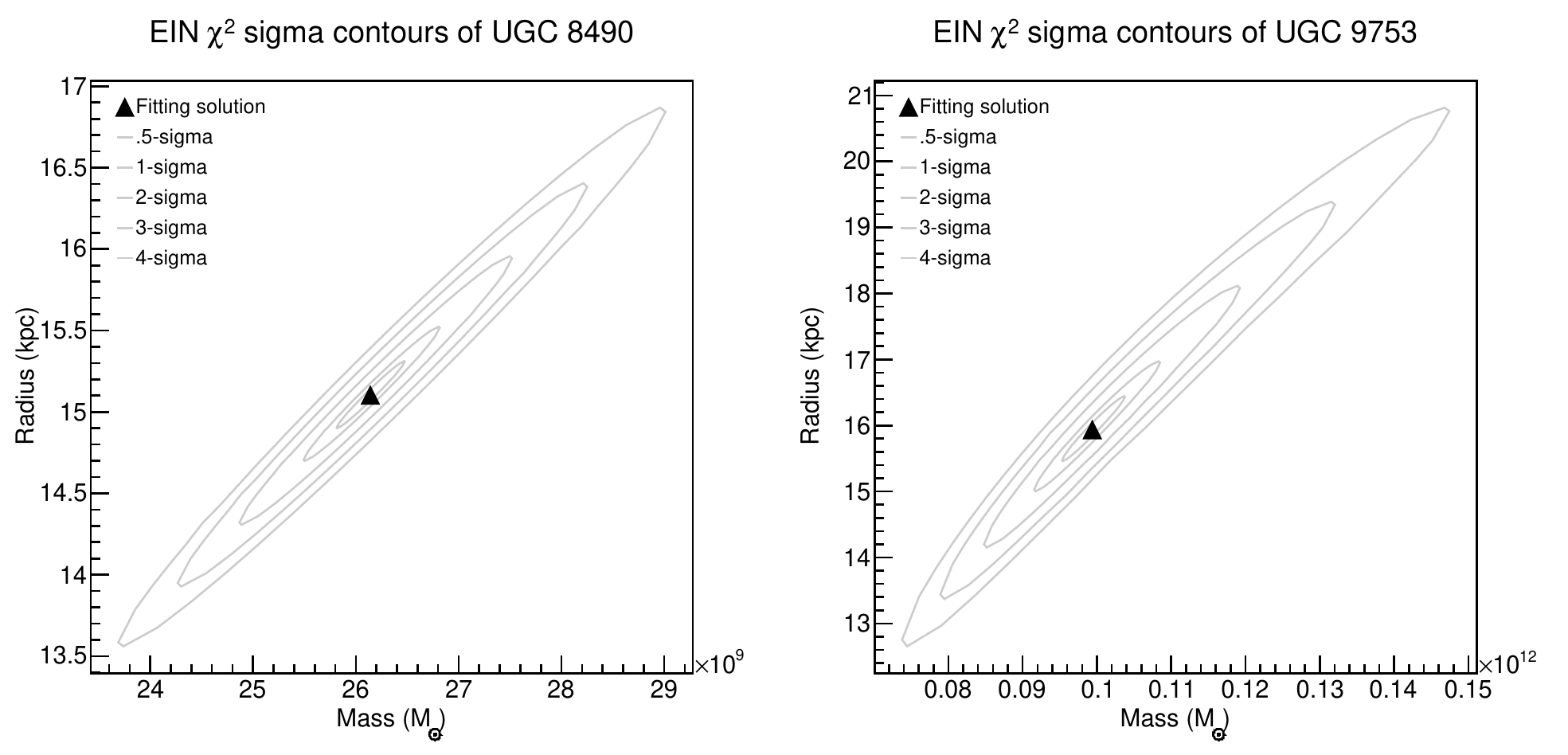}
\caption[f8.pdf]{{\it{Left}}: EIN $\chi^2$ sigma contours of the EIN spherical DM fit to the HI DM RC of UGC 8490. The description of the graphic content is identical to that of figure~(\ref{fig6}). {\it{Right}}: Same information for UGC 9753.}
~\label{fig8} 
\end{figure*}

\begin{figure*}
\centering
\includegraphics[width=1.0\hsize]{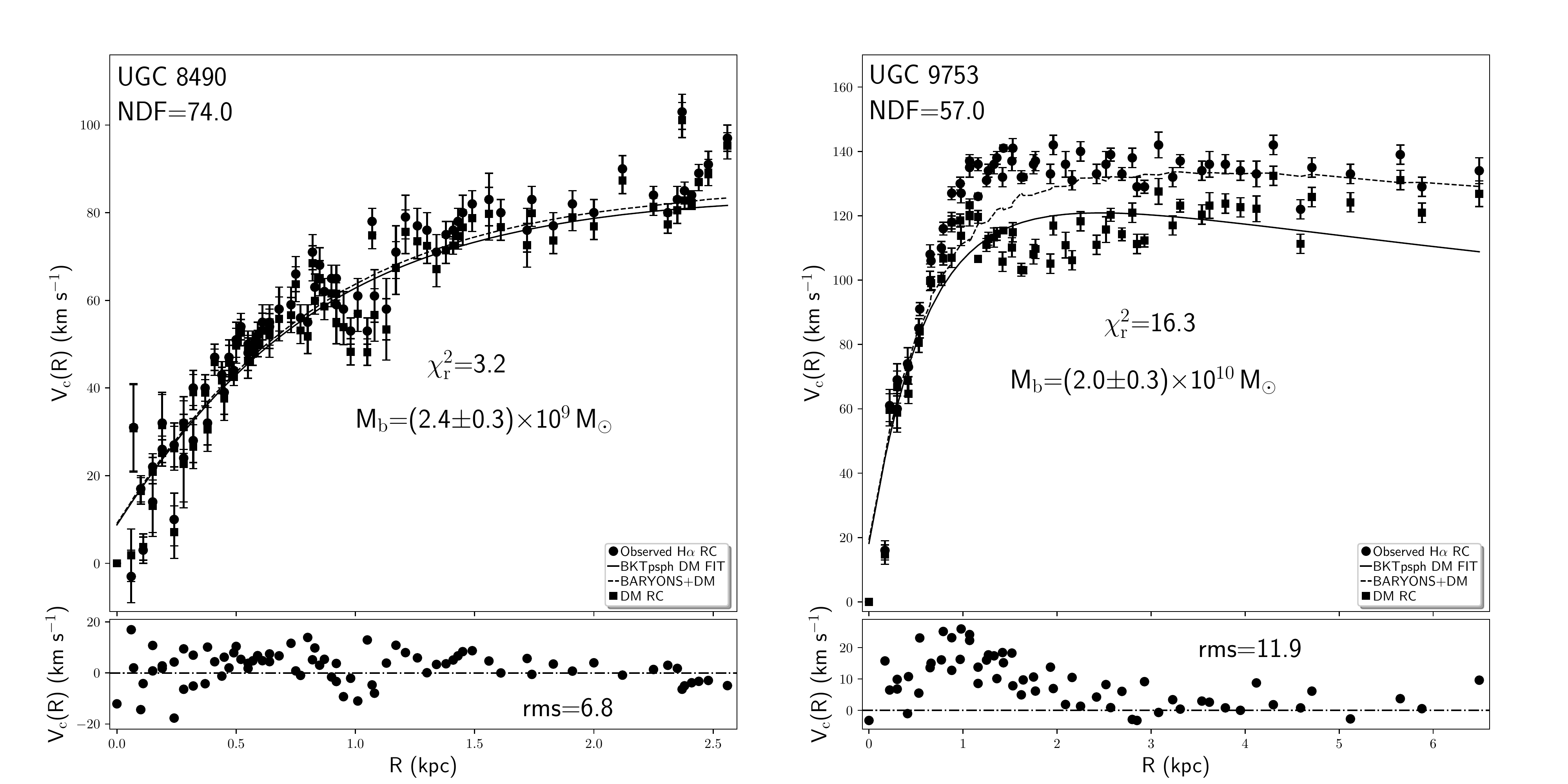}
\caption[f9.pdf]{{\it{Left}}: BKT prolate spheroidal fit to the H$\alpha$ DM RC of UGC 8490. The description of the graphic content is identical to that of figure~(\ref{fig5}). {\it{Right}}: Same information for UGC 9753.}
~\label{fig9}
\end{figure*}

\begin{figure*}
\centering
\includegraphics[width=1.0\hsize]{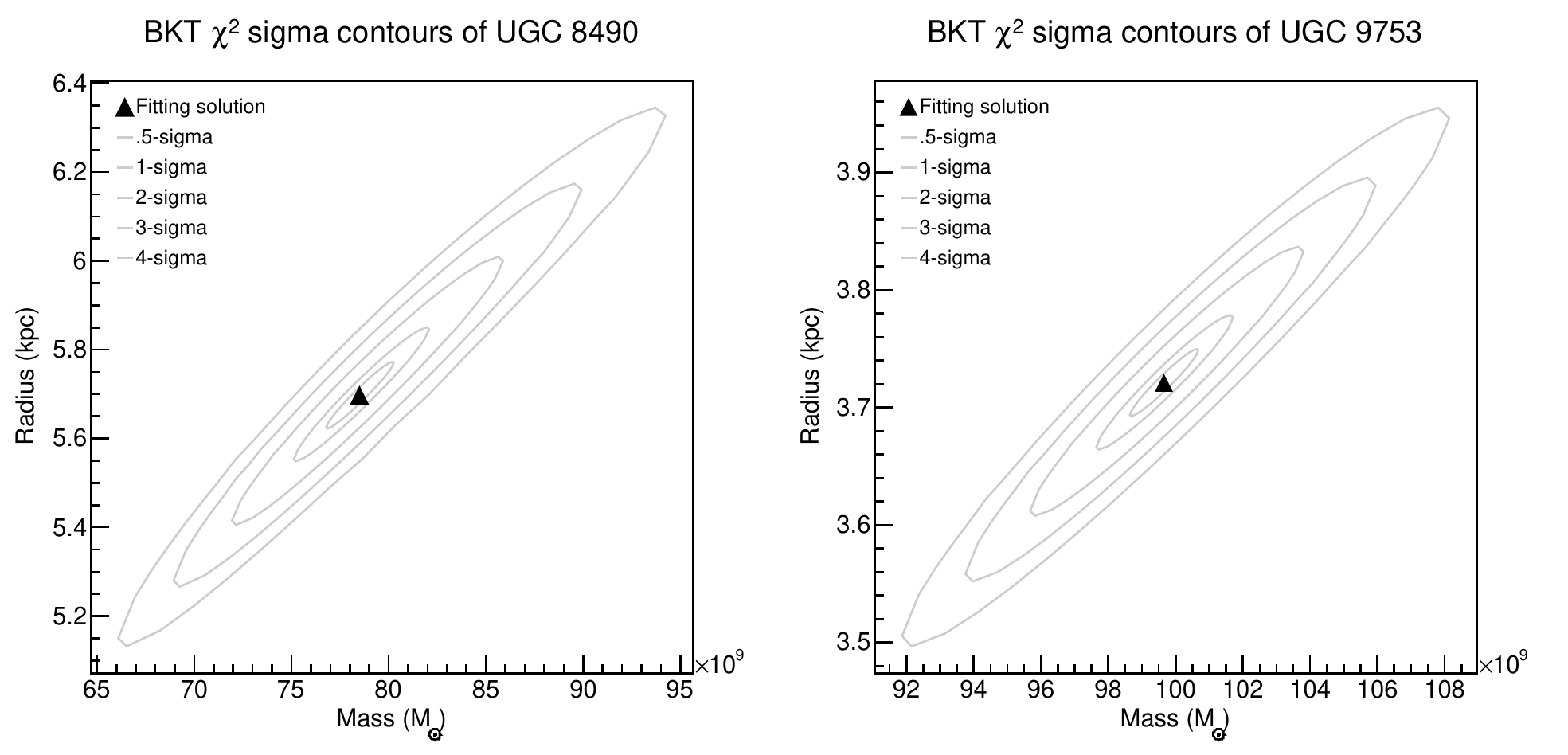}
\caption[f10.pdf]{{\it{Left}}: BKT $\chi^2$ sigma contours of the BKT prolate spheroidal DM fit to the H$\alpha$ DM RC of UGC 8490. The description of the graphic content is identical to that of figure~(\ref{fig6}). {\it{Right}}: Same information for UGC 9753.}
~\label{fig10} 
\end{figure*}

\begin{figure*}
\centering
\includegraphics[width=1.0\hsize]{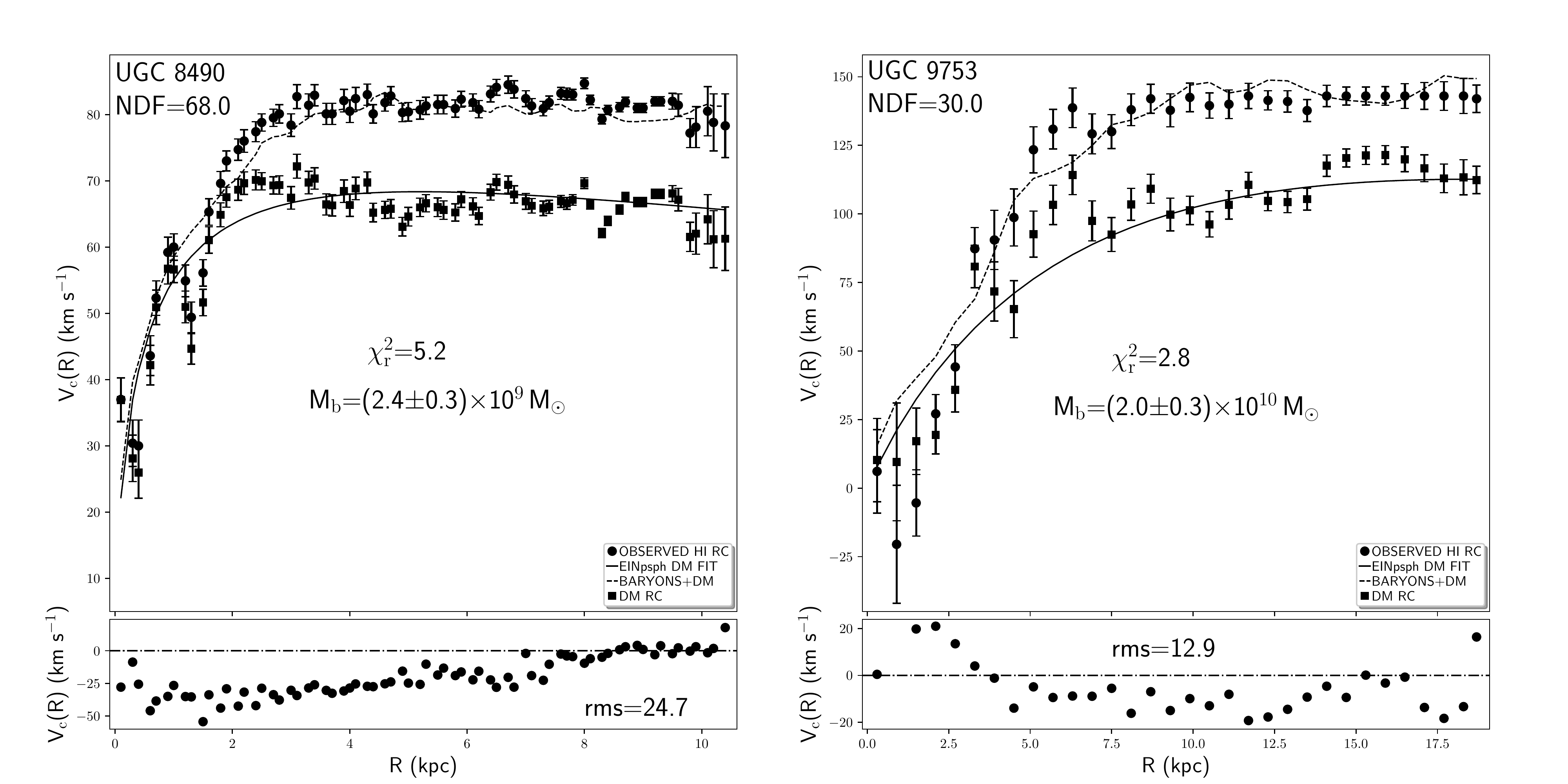}
\caption[f11.pdf]{{\it{Left}}: EIN prolate spheroidal fit to the HI DM RC of UGC 8490. The description of the graphic content is identical to that of figure~(\ref{fig7}). {\it{Right}}: Same information for UGC 9753.}
~\label{fig11}
\end{figure*}

\begin{figure*}
\centering
\includegraphics[width=1.0\hsize]{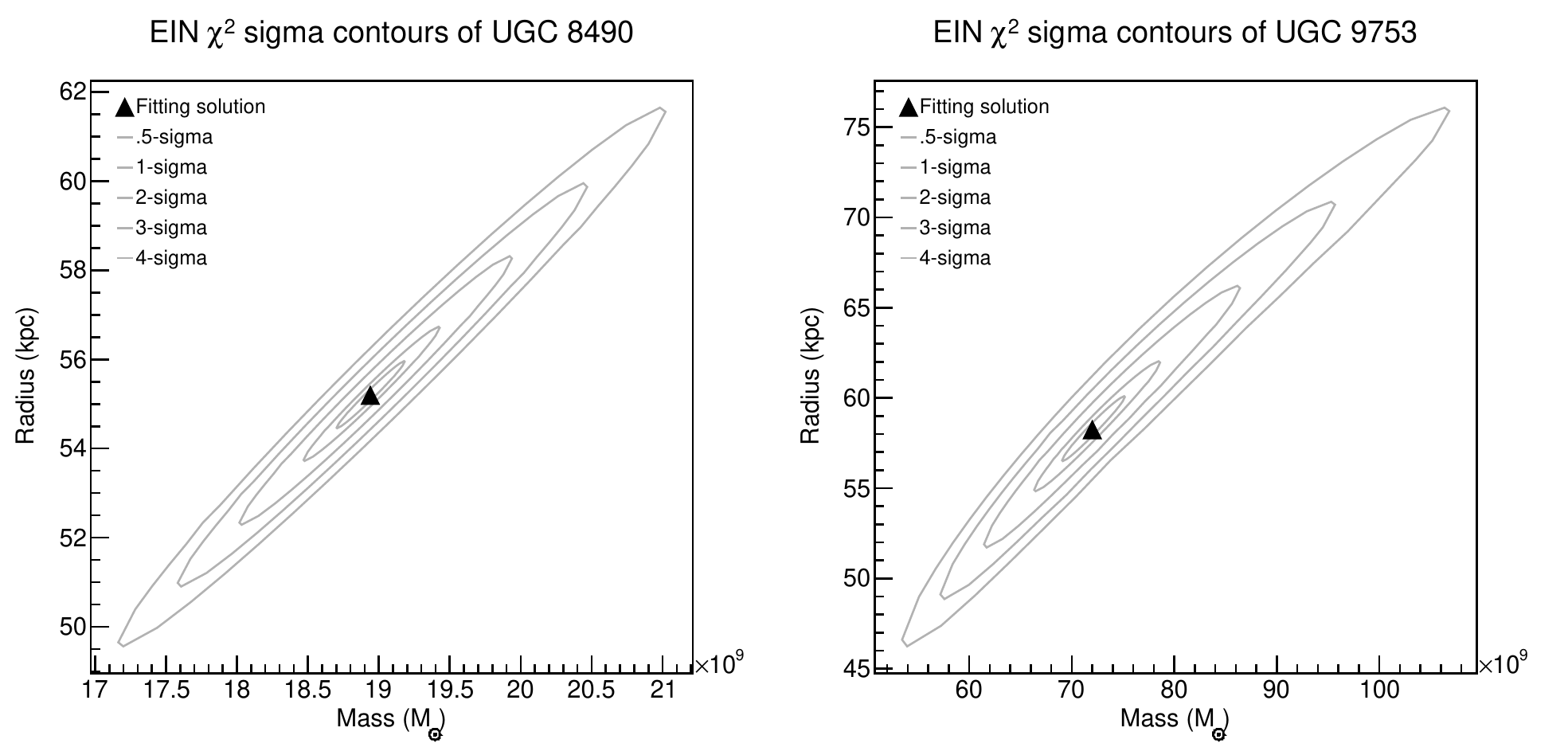}
\caption[f12.pdf]{{\it{Left}}: EIN $\chi^2$ sigma contours of the EIN prolate spheroidal DM fit to the HI DM RC of UGC 8490. The description of the graphic content is identical to that of figure~(\ref{fig8}). {\it{Right}}: Same information for UGC 9753.}
~\label{fig12} 
\end{figure*}

\begin{figure*}
\centering
\includegraphics[width=1.0\hsize]{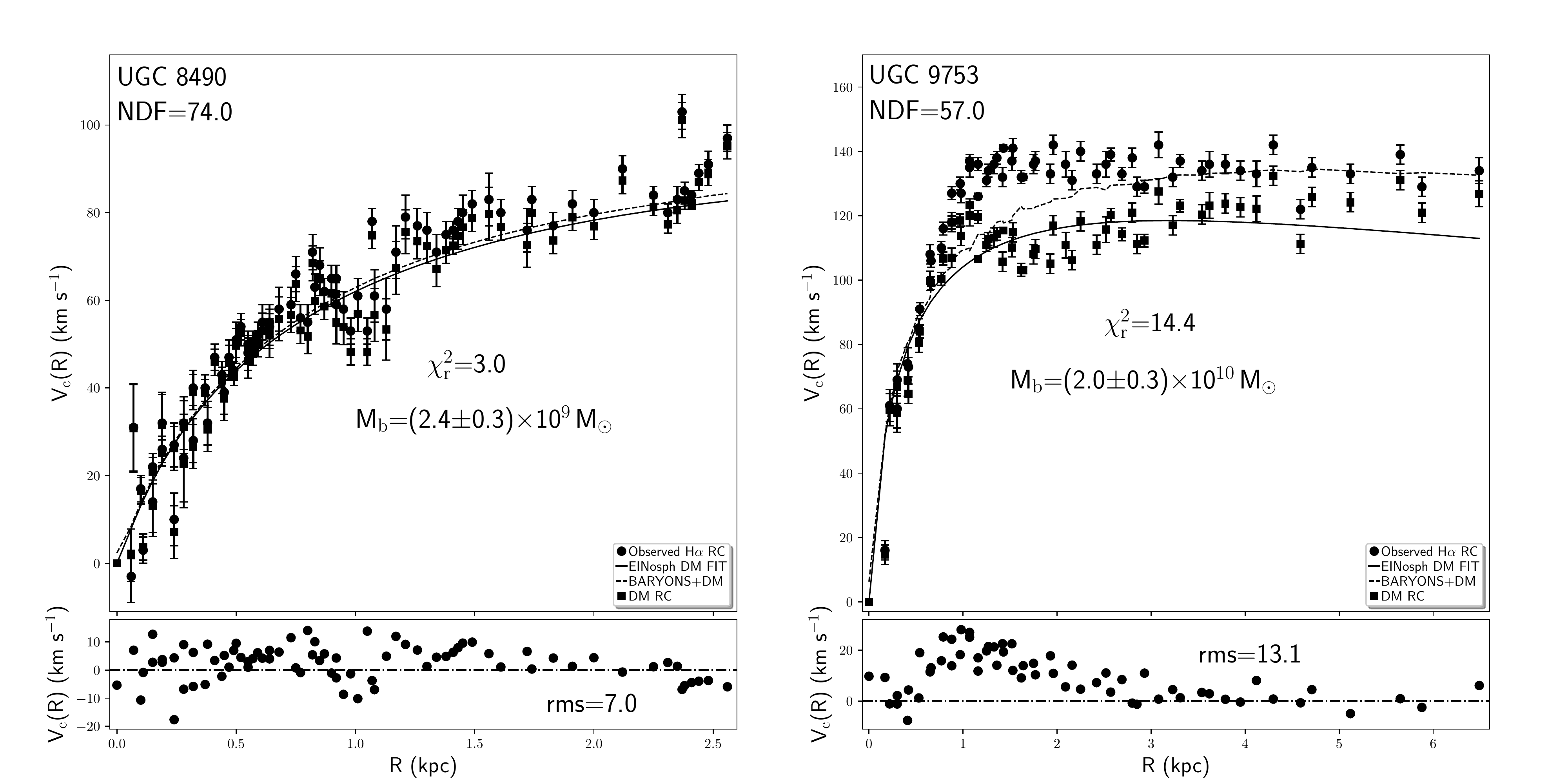}
\caption[f13.pdf]{{\it{Left}}: EIN oblate spheroidal fit to the H$\alpha$ DM RC of UGC 8490. The description of the graphic content is identical to that of figure~(\ref{fig5}). {\it{Right}}: Same information for UGC 9753.}
~\label{fig13}
\end{figure*}

\begin{figure*}
\centering
\includegraphics[width=1.0\hsize]{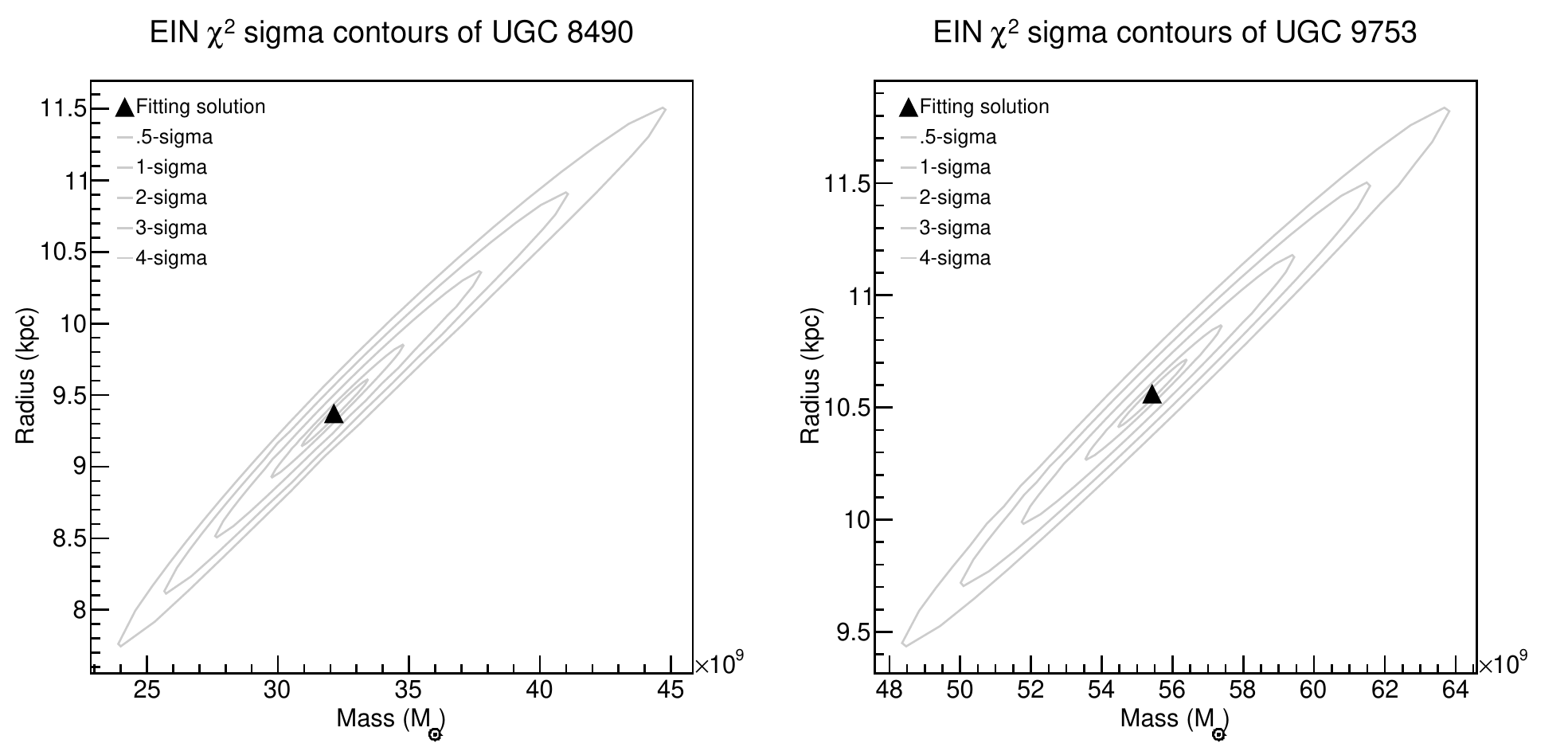}
\caption[f14.pdf]{{\it{Left}}: EIN $\chi^2$ sigma contours of the EIN oblate spheroidal DM fit to the H$\alpha$ DM RC of UGC 8490. The description of the graphic content is identical to that of figure~(\ref{fig6}). {\it{Right}}: Same information for UGC 9753.}
~\label{fig14} 
\end{figure*}

\begin{figure*}
\centering
\includegraphics[width=1.0\hsize]{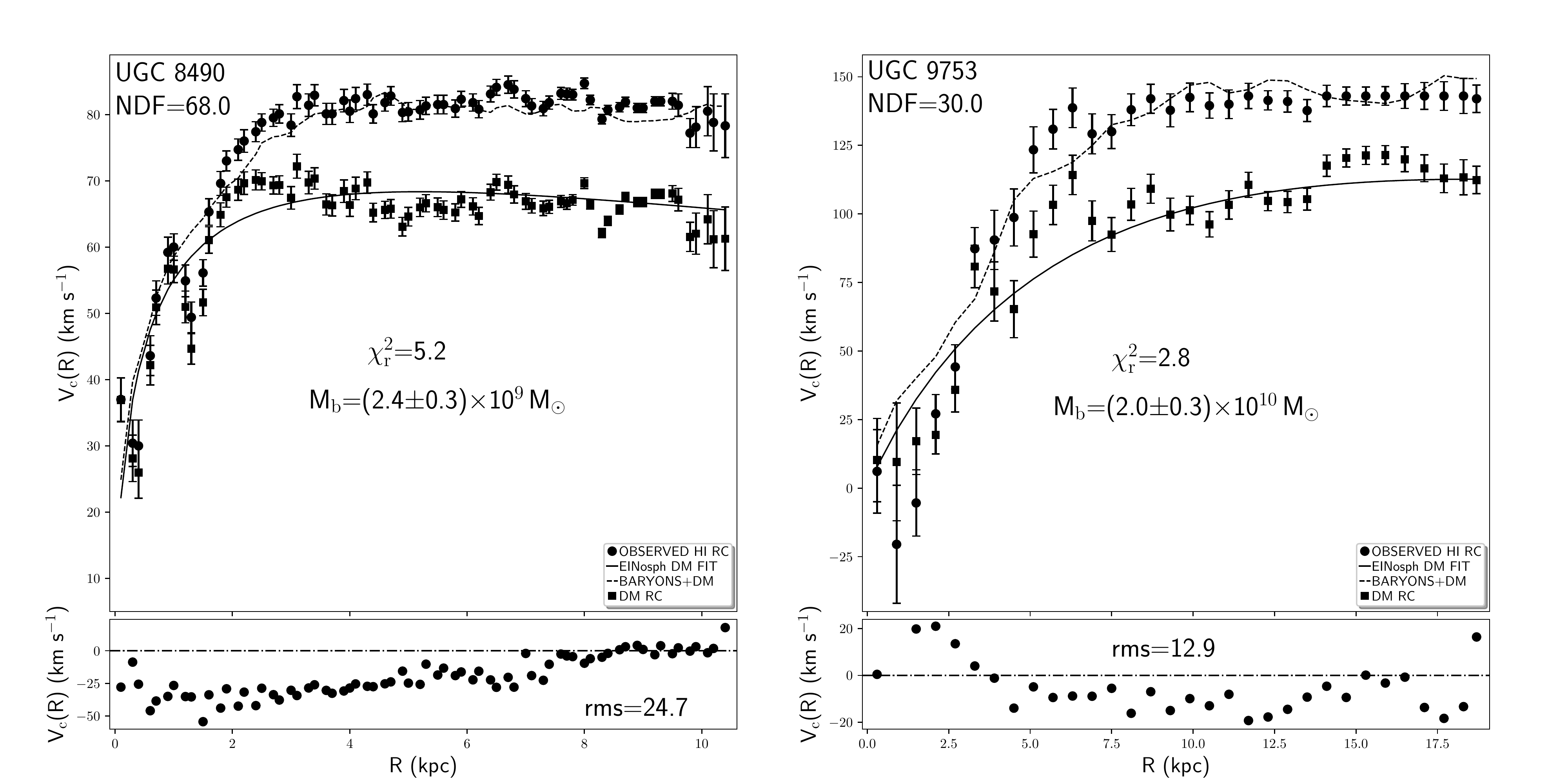}
\caption[f15.pdf]{{\it{Left}}: EIN oblate spheroidal fit to the HI DM RC of UGC 8490. The description of the graphic content is identical to that of figure~(\ref{fig7}). {\it{Right}}: Same information for UGC 9753.}
~\label{fig15}
\end{figure*}

\begin{figure*}
\centering
\includegraphics[width=1.0\hsize]{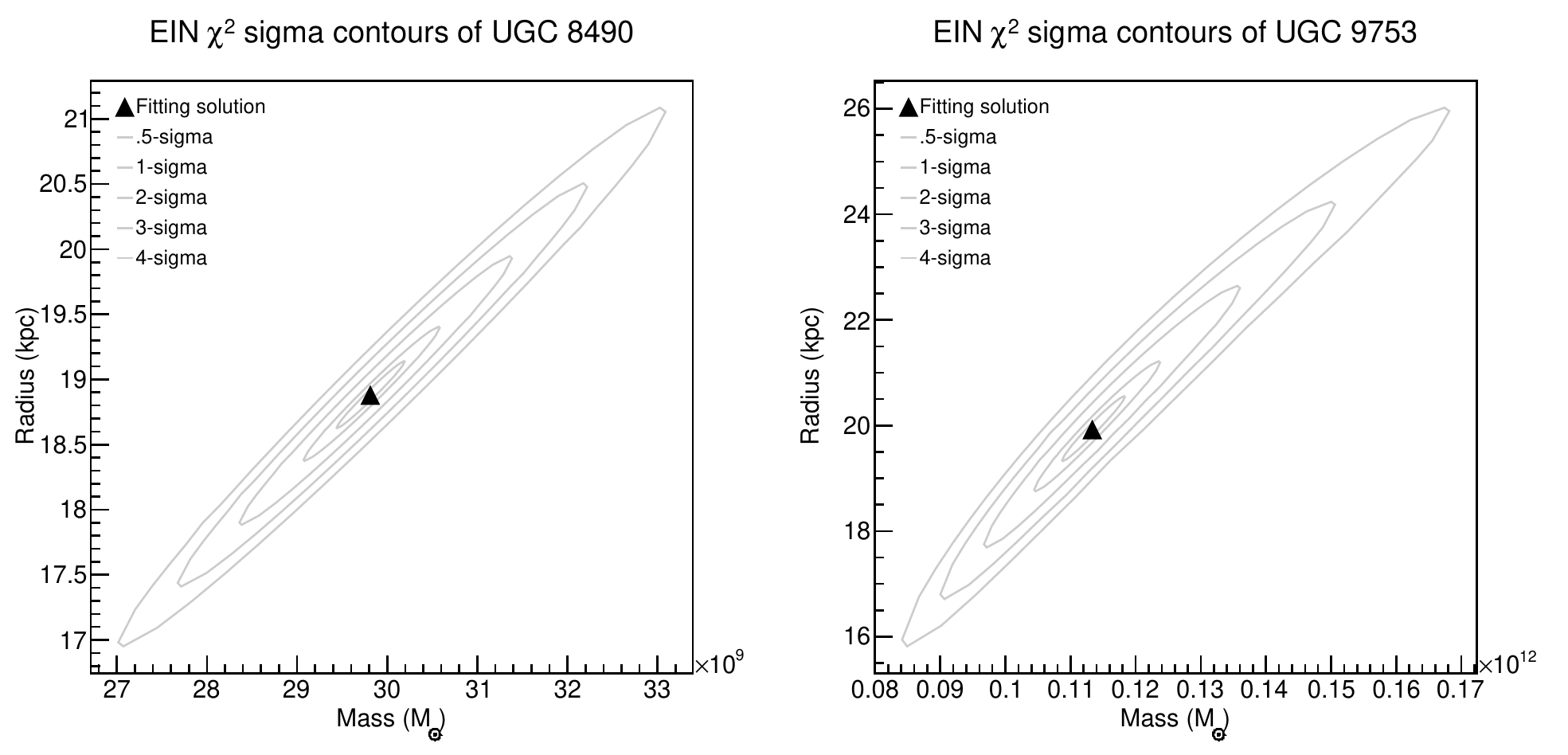}
\caption[f16.pdf]{{\it{Left}}: EIN $\chi^2$ sigma contours of the EIN oblate spheroidal DM fit to the HI DM RC of UGC 8490. The description of the graphic content is identical to that of figure~(\ref{fig8}). {\it{Right}}: Same information for UGC 9753.}
~\label{fig16} 
\end{figure*}

In this section we outline the principal results that are also summarised in the corresponding tables and figures. For the spherical, prolate and oblate spheroidal mass distributions and the H$\alpha$ DM RCs fit of UGC 8490 and UGC 9753, the NFW spherical solution is better than the respective prolate and oblate solutions for both galaxies, whereas the prolate BKT performs better than the oblate and spherical BKT. The DCN spherical result is the best for UGC 8490, on the other hand the oblate DCN adjusts better the H$\alpha$ RC of UGC 9753. The oblate EIN fits better than the spherical and prolate EIN the H$\alpha$ DM RCs of both galaxies. The STD prolate fit to the H$\alpha$ DM RC of UGC 8490 is the best, while on the contrary, the spherical STD adjusts better the H$\alpha$ DM RC of UGC 9753. For the HI DM RC of both galaxies the prolate NFW is the best. The spherical, prolate and oblate BKT fits to the HI DM RCs of both galaxies are identical. The prolate DCN performs better than the spherical and oblate DCN fits. The oblate EIN is the best for the HI DM RCs of both galaxies. The spherical STD is the best for the HI DM RCs of both objects. In general cored DM halos EIN and BKT predominate over cuspy NFW and DCN. In particular EIN cored solutions are the best for the H$\alpha$ DM RCs of both galaxies as well as for the HI DM RC of UGC 8490, whereas the cored BKT is the best for the HI DM RC of UGC 9753. The STD cored DM halo solutions are comparable to the EIN and BKT solutions, however these results do not perform better than the EIN and BKT DM halos for all the DM RCs data used in this analysis. In general the NFW and DCN behave in a similar manner, nevertheless the spherical, prolate and oblate NFW DM halo perform better than the DCN DM halo, except for the prolate NFW fit to the H$\alpha$ DM RC of UGC 8490, where the DCN DM halo provides a better result. In brief, the main results are that, cored spherical/oblate EIN and BKT fit better than any other DM halos considered in this study the DM RCs of UGC 8490 and UGC 9753.
The resultant masses and scale radii together with other relevant information are tabulated in tables~(\ref{tab:tb3}),~(\ref{tab:tb4}) and~(\ref{tab:tb5}), and the best spherical, prolate and oblate fits to the H$\alpha$ and HI DM RCs are displayed from figures~(\ref{fig5}) to~(\ref{fig16}). 

\begin{table}
\centering
\caption{DM and total mass of UGC 8490 and UGC 9753: spherical mass distribution H$\alpha$ and HI DM RCs}
\label{tab:tb3}
\begin{threeparttable}
\begin{tabular}{p{0.9cm}p{0.8cm}p{0.9cm}p{1.0cm}p{0.5cm}p{0.6cm}}
\hline
DMLs\tnote{(1)} & M$_T$\tnote{(2)}  & M$_{DM}$\tnote{(3)}  & R$_h$\tnote{(4)} & $\chi^2_r$\tnote{(5)} & EDM\tnote{(6)}\\
\hline
NFW & 2.4$\pm$0.1 & 2.2$\pm$0.1 & 15.2$\pm$0.8 & 3.1 & 9.6$\times$10$^{-9}$\\ 

BKT & 0.3$\pm$0.04 & 0.1$\pm$0.01 & 0.9$\pm$0.02 & 3.5 & 9.9$\times$10$^{-8}$\\

DCN & 3.0$\pm$0.1 & 2.8$\pm$0.1 & 5.4$\pm$0.1 & 3.1 & 6.5$\times$10$^{-7}$\\

EIN & 23.0$\pm$3.0 & 21.0$\pm$3.0 & 83.5$\pm$9.1 & 3.0 & 5.0$\times$10$^{-10}$\\

STD & 4.2$\pm$0.2 & 4.0$\pm$0.2 & 4.6$\pm$0.2 & 9.5 & 4.2$\times$10$^{-9}$\\
\hline
NFW & 4.0$\pm$0.4 & 2.0$\pm$0.1 & 6.4$\pm$0.2 & 14.8 & 7.8$\times$10$^{-7}$\\ 

BKT & 2.1$\pm$0.3 & 0.1$\pm$0.03 & 0.7$\pm$0.01 & 16.7 & 5.5$\times$10$^{-9}$\\

DCN &  6.1$\pm$0.3 & 4.1$\pm$0.01 & 3.2$\pm$0.04 & 28.4 & 6.3$\times$10$^{-10}$\\

EIN &  12.0$\pm$0.8 & 9.9$\pm$0.5 & 27.0$\pm$1.5 & 14.4 & 6.9$\times$10$^{-9}$\\

STD & 47.0$\pm$4.0 & 45.0$\pm$4.0 & 0.34$\pm$0.03 & 15.2 & 1.1$\times$10$^{-6}$\\
\hline
\hline
NFW & 1.5$\pm$0.06 & 1.3$\pm$0.03 & 14.7$\pm$0.4 & 6.0 & 3.6$\times$10$^{-7}$\\ 

BKT & 0.3$\pm$0.04 & 0.1$\pm$0.01 & 1.2$\pm$0.02 & 7.5 & 1.0$\times$10$^{-9}$\\

DCN & 3.0$\pm$0.06 & 2.8$\pm$0.03 & 11.0$\pm$0.1 & 6.6 & 3.4$\times$10$^{-10}$\\

EIN & 2.8$\pm$0.1 & 2.6$\pm$0.07 & 15.1$\pm$0.4 & 5.2 & 3.4$\times$10$^{-7}$\\

STD & 3.0$\pm$0.2 & 2.8$\pm$0.2 & 1.2$\pm$0.08 & 5.2 & 2.3$\times$10$^{-9}$\\
\hline
NFW & 25.0$\pm$0.7 & 23.0$\pm$0.4 & 58.0$\pm$1.3 & 3.7 & 1.9$\times$10$^{-7}$\\ 

BKT & 3.1$\pm$0.4 & 1.1$\pm$0.1 & 6.3$\pm$0.4 & 2.7 & 2.0$\times$10$^{-8}$\\

DCN & 33.1$\pm$3.8 & 31.1$\pm$3.5 & 42.1$\pm$4.8 & 5.5 & 5.8$\times$10$^{-12}$\\

EIN & 12.0$\pm$1.1 & 9.9$\pm$0.8 & 16.0$\pm$0.1 & 2.8 & 5.3$\times$10$^{-8}$\\

STD & 27.4$\pm$2.0 & 25.4$\pm$1.7 & 20.2$\pm$1.1 & 2.7 & 4.7$\times$10$^{-7}$\\
\hline
\end{tabular}
\begin{tablenotes}
\item[1] DM density laws.
\item[2] Total mass in units of 10$^{10}$ (M$_{\odot}$).
\item[3] DM mass in units of 10$^{10}$ (M$_{\odot}$).
\item[4] DM scale radius (kpc).
\item[5] Reduced $\chi^2$.
\item[6] Estimated distance to the minimum.
\end{tablenotes}
\end{threeparttable}
\end{table}


\begin{table}
\centering
\caption{DM and total mass of UGC 8490 and UGC 9753: prolate spheroidal mass distribution H$\alpha$ and HI DM RCs}
\label{tab:tb4}
\begin{threeparttable}
\begin{tabular}{p{0.8cm}p{0.8cm}p{0.9cm}p{0.8cm}p{0.2cm}p{0.2cm}p{0.3cm}p{0.3cm}}
\hline
DMLs\tnote{(1)} & M$_T$\tnote{(2)}  & M$_{DM}$\tnote{(3)}  & R$_h$\tnote{(4)} & a\tnote{(5)} & b\tnote{(6)} & $\chi^2_r$\tnote{(7)} & EDM\tnote{(8)}\\
\hline
NFW &  30.2$\pm$0.4 & 30.0$\pm$0.4 & 65.2$\pm$0.2 & 0.4 & 0.7 & 7.2 & 1.4$\times$10$^{-8}$\\ 

BKT & 8.0$\pm$0.4 & 7.8$\pm$0.4 & 5.7$\pm$0.2 & 0.4 & 0.7 & 3.2 & 1.5$\times$10$^{-9}$\\

DCN & 11.2$\pm$0.8 & 11.0$\pm$0.8 & 24.2$\pm$1.1 & 0.8 & 1.5 & 6.5 & 2.2$\times$10$^{-8}$\\

EIN & 0.7$\pm$0.3 & 0.5$\pm$0.02 & 10.1$\pm$0.3 & 0.4 & 0.7 & 3.3 & 7.6$\times$10$^{-8}$\\

STD & 69.2$\pm$3.0 & 69.0$\pm$3.0 & 19.7$\pm$0.6 & 0.4 & 0.7 & 3.1 & 1.0$\times$10$^{-9}$\\
\hline
NFW &  32.4$\pm$1.3 & 30.4$\pm$1.0 & 54.4$\pm$1.3 & 0.3 & 0.9 & 17.2 & 5.5$\times$10$^{-10}$\\ 

BKT & 12.0$\times$0.5 & 10.0$\times$0.2 & 3.7$\times$0.06 & 0.2 & 0.6 & 16.3 & 1.7$\times$10$^{-9}$\\

DCN & 2.3$\pm$0.3 & 0.3$\pm$0.01 & 1.5$\pm$0.08 & 0.5 & 1.6 & 22.6 & 3.6$\times$10$^{-9}$\\

EIN & 4.0$\pm$0.4  & 2.0$\pm$0.06 & 20.4$\pm$0.5 & 0.3 & 0.8 & 16.8 & 1.3$\times$10$^{-11}$\\

STD & 81.0$\pm$1.7 & 79.0$\pm$1.4 & 48.7$\pm$1.6 & 0.2 & 0.6 & 21.0 & 3.4$\times$10$^{-9}$\\
\hline
\hline  
NFW & 17.3$\pm$0.4 & 17.1$\pm$0.4 & 78.0$\pm$1.8 & 0.4 & 0.7 & 5.1 & 1.5$\times$10$^{-7}$\\ 

BKT & 1.8$\pm$0.05 & 1.6$\pm$0.02 & 3.9$\pm0.07$ & 0.4 & 0.8 & 7.5 & 6.4$\times$10$^{-9}$\\

DCN & 55.0$\pm$1.2 & 53.0$\pm$1.2 & 56.0$\pm$1.7 & 0.4 & 0.8 & 6.1 & 3.7$\times$10$^{-9}$\\

EIN &  2.1$\pm$0.08 & 1.9$\pm$0.05 & 55.2$\pm$1.5 & 0.5 & 0.9 & 5.2 & 2.1$\times$10$^{-11}$\\

STD & 92.8$\pm$1.2 & 92.6$\pm$1.2 & 57.0$\pm$0.9 & 0.4 & 0.8 & 7.8 & 1.3$\times$10$^{-7}$\\
\hline
NFW & 75.0$\pm$2.3 & 73.0$\pm$2.0 & 97.1$\pm$2.3 & 0.3 & 0.8 & 3.7 & 2.5$\times$10$^{-7}$\\ 

BKT & 25.1$\pm$2.3 & 23.1$\pm$2.0 & 20.3$\pm$1.3 & 0.3 & 0.9 & 2.7 & 1.7$\times$10$^{-8}$\\

DCN &  2.2$\pm$0.3 & 0.2$\pm$0.02 & 2.1$\pm$0.2 & 0.2 & 0.7 & 4.2 & 3.9$\times$10$^{-9}$\\

EIN & 9.2$\pm$0.9 & 7.2$\pm$0.6 & 58.2$\pm$3.6 & 0.3 & 0.8 & 2.8 & 4.1$\times$10$^{-10}$\\

STD & 382$\pm$16 & 380$\pm$15.7 & 84.0$\pm$2.5 & 0.23 & 0.7 & 2.9 & 3.7$\times$10$^{-7}$\\
\hline
\end{tabular}
\begin{tablenotes}
\item[1] DM density laws.
\item[2] Total mass in units of 10$^{10}$ (M$_{\odot}$).
\item[3] DM mass in units of 10$^{10}$ (M$_{\odot}$).
\item[4] DM scale radius (kpc).
\item[5] Prolate spheroid minor axis.
\item[6] Prolate spheroid major axis.
\item[7] Reduced $\chi^2$.
\item[8] Estimated distance to the minimum.
\end{tablenotes}
\end{threeparttable}
\end{table}


\begin{table}
\centering
\caption{DM and total mass of UGC 8490 and UGC 9753: oblate spheroidal mass distribution H$\alpha$ and HI DM RCs}
\label{tab:tb5}
\begin{threeparttable}
\begin{tabular}{p{0.8cm}p{0.8cm}p{0.9cm}p{0.8cm}p{0.2cm}p{0.2cm}p{0.3cm}p{0.3cm}}
\hline
DMLs\tnote{(1)} & M$_T$\tnote{(2)}  & M$_{DM}$\tnote{(3)}  & R$_h$\tnote{(4)} & a\tnote{(5)} & b\tnote{(6)} & $\chi^2_r$\tnote{(7)} & EDM\tnote{(8)}\\
\hline
NFW &  54.4$\pm$0.7 & 54.2$\pm$0.7 & 63.4$\pm$2.0 & 0.6 & 0.7 & 4.6 & 9.2$\times$10$^{-10}$\\ 

BKT & 1.4$\pm$0.08 & 1.2$\pm$0.05 & 1.7$\pm$0.04 & 0.7 & 0.8 & 3.2 & 5.4$\times$10$^{-9}$\\

DCN & 9.9$\pm$0.8 & 9.7$\pm$0.8 & 23.0$\pm$1.1 & 2.1 & 2.4 & 5.9 & 3.2$\times$10$^{-7}$\\

EIN & 3.4$\pm$0.3 & 3.2$\pm$0.3 & 9.4$\pm$0.5 & 0.6 & 0.7 & 3.0 & 3.0$\times$10$^{-8}$\\

STD & 6.9$\pm$0.1 & 6.7$\pm$0.1 & 3.0$\pm$0.04 & 0.7 & 0.8 & 5.1 & 2.5$\times$10$^{-9}$\\
\hline
NFW & 13.1$\pm$0.7 & 11.1$\pm$0.4 & 23.1$\pm$0.6 & 0.7 & 0.8 & 17.2 & 1.8$\times$10$^{-7}$\\ 

BKT & 3.5$\pm$0.3 & 1.5$\pm$0.03 & 1.1$\pm$0.02 & 0.6 & 0.7 & 16.3 & 5.4$\times$10$^{-8}$\\

DCN & 2.3$\pm$0.3 & 0.3$\pm$0.03 & 2.5$\pm$0.03 & 2.0 & 2.3 & 22.6 & 3.5$\times$10$^{-12}$\\

EIN & 7.5$\pm$0.5 & 5.5$\pm$0.2 & 11.0$\pm$0.3 & 0.7 & 0.8 & 14.4 & 5.2$\times$10$^{-10}$\\

STD & 111$\pm$4.0 & 109$\pm$3.7 & 18.3$\pm$0.6 & 0.6 & 0.7 & 20.0 & 1.9$\times$10$^{-10}$\\
\hline
\hline
NFW & 3.4$\pm$0.4 & 3.2$\pm$0.07 & 28.0$\pm$0.6 & 0.6 & 0.7 & 5.1 & 6.2$\times$10$^{-8}$\\ 

BKT & 0.8$\pm$0.1 & 0.6$\pm$0.08 & 1.5$\pm$0.03 & 0.6 & 0.7 & 7.5 & 3.0$\times$10$^{-11}$\\

DCN & 7.2$\pm$0.1 & 7.0$\pm$0.08 & 11.2$\pm$0.1 & 0.8 & 0.9 & 7.9 & 9.2$\times$10$^{-8}$\\

EIN & 3.2$\pm$0.1 & 3.0$\pm$0.08 & 19.0$\pm$0.5 & 0.6 & 0.7 & 5.2 & 2.1$\times$10$^{-10}$\\

STD & 21.2$\pm$0.3 & 21.0$\pm$0.3 & 19.0$\pm$0.3 & 0.7 & 0.8 & 7.8 & 2.8$\times$10$^{-7}$\\
\hline
NFW & 37.1$\pm$0.9 & 35.1$\pm$0.6 & 62.3$\pm$0.7 & 0.8 & 0.9 & 5.3 & 3.4$\times$10$^{-7}$\\ 

BKT & 10.0$\pm$1.0 & 8.0$\pm$0.7 & 7.8$\pm$0.5 & 0.7 & 0.8 & 2.7 & 6.1$\times$10$^{-9}$\\

DCN &  2.2$\pm$0.3 & 0.2$\pm$0.02 & 1.8$\pm$0.1 & 0.7 & 0.8 & 4.2 & 2.2$\times$10$^{-9}$\\

EIN & 13.3$\pm$1.3 & 11.3$\pm$1.0 & 20.0$\pm$1.2 & 0.7 & 0.8 & 2.8 & 1.1$\times$10$^{-8}$\\

STD & 101$\pm$7.7 & 99.0$\pm$7.4 & 32.0$\pm$1.7 & 0.6 & 0.7 & 2.8 & 4.0$\times$10$^{-8}$\\
\hline
\end{tabular}
\begin{tablenotes}
\item[1] DM density laws.
\item[2] Total mass in units of 10$^{10}$ (M$_{\odot}$).
\item[3] DM mass in units of 10$^{10}$ (M$_{\odot}$).
\item[4] DM scale radius (kpc).
\item[5] oblate spheroid minor axis.
\item[6] oblate spheroid major axis.
\item[7] Reduced $\chi^2$.
\item[8] Estimated distance to the minimum.
\end{tablenotes}
\end{threeparttable}
\end{table}

\subsection{Comparison with other works}\label{sec:s11}

In this section we compare the results obtained for UGC 8490 and UGC 9753 with the findings of other authors that studied the DM and total mass distribution of the same galaxies through the fit of the H$\alpha$ and HI RCs. In the following we contrast our results only with the findings of those studies that have considered the DM (see also section~(\ref{sec:s13})) as the principal counterpart of the stellar and gaseous components. We outline the most important findings of the comparable studies below, whereas we provide additional parameters and information in table~(\ref{tab:tb6}). \citet{Sicotte1996} and \citet{Sicotte1997} studied, among other things, the H$\alpha$ and HI kinematic and mass content of UGC 8490 determining that the DM forms 86$\%$ of the entire gravitating mass. The DM dominates over the baryonic component at a radius of 1.7 kpc from the center of the object. In addition the authors found that the DM mass distribution of UGC 8490 is well represented by a pseudo-isothermal (PISO) DM halo. \citet{vandenBosch2001} analised the DM slopes of UGC 8490 together with other 19 dwarf galaxies, establishing that the DM halo of UGC 8490 is consistent with a NFW density profile and determined the corresponding concentration and circular velocity at a density contrast of 200. \citet{Swaters2003} examined the mass content of UGC 8490 using H$\alpha$ and HI RCs and found that its DM distribution is compatible with a PISO DM halo. \citet{Spano2008} derived composite H$\alpha$ plus HI RCs of 36 spiral galaxies that include UGC 8490 to analyse the mass content of these objects. The authors found that the DM distribution of UGC 8490 is concordant with a PISO DM profile. \citet{Jimenez2003} analysed the DM distribution of 400 galaxies of different morphological types, using optical H$\alpha$ RCs from the literature. The authors considered the NFW and the PISO DM halos to fit the observed RCs in order to obtain the DM halo parameters and to determine the central density slope of the studied galaxies. 

\citet{Jimenez2003} results for UGC 9753 indicate that this object has a finite central density, and the best fitted model is represented by the PISO DM halo. We are not aware of other research works that examine the total mass content or the DM distribution of UGC 9753. We have to emphasise that all the studies we have mentioned in this section assume an spherical DM distribution. As a consequence (see table~\ref{tab:tb6}) we only compare the findings generated by the spherical fits to the H$\alpha$ and HI DM RCs with the corresponding results of the above cited authors. As it is clear from table~(\ref{tab:tb6}) our spherical results are in relatively good agreement with the corresponding findings of the works reported in the same table. The differences in the mass values are probably due to the diverse DM halos used (PISO versus BKT), the distinct H$\alpha$ and HI RCs utilized by the other studies, the fact that in all the research works considered for comparison, the fits were done as the combination of the H$\alpha$ and HI RCs, whereas in the present analysis we adjusted separately the H$\alpha$ and HI DM RCs, the dissimilar derivation of the stellar and gas RCs that in our specific case are not treated as parametric components, and ultimately the fact that we fit the DM RC and not the total observed RC. The discrepancy in the values of concentration and circular velocity at the density contrast 200 is likely imputable to the different definition of the concentration parameter. In this work we use the recipe of \citet{Dutton2014}, whereas the study of \citet{vandenBosch2001} utilized the original prescription of \citet{Navarro1996}. The differences are also due to the diverse RCs they used and to the distinct method employed to account for the baryonic part of UGC 8490. In general, regardless of all these differences, the comparison denotes that our results are consistent with the findings of other studies.             

\begin{table}
\centering
\caption{Comparison of the spherical fits results with those of other authors for the H$\alpha$ and HI DM RCs of UGC 8490.}
\label{tab:tb6}
\begin{threeparttable}
\begin{tabular}{p{1.6cm}p{1.3cm}p{1.3cm}p{1.3cm}}
\hline
M$_h$\tnote{(1)} & r$_h$\tnote{(2)}  & c$_{200}$\tnote{(3)} & V$_{200}$\tnote{(4)}\\
\hline
3.4$\times$10$^9$\tnote{(5)} & 2.7\tnote{(5)} & --- & ---\\
--- & --- & 13.5\tnote{(6)} & 57.2\tnote{(6)}\\
3.1$\times$10$^8$\tnote{(7)} & 0.56\tnote{(7)} & --- & ---\\
2.7$\times$10$^9$\tnote{(8)} & 1.9\tnote{(8)} & --- & ---\\
1.0$\times$10$^9$\tnote{(9)} & 0.9\tnote{(9)} & --- & ---\\
1.0$\times$10$^9$\tnote{(9)} & 1.2\tnote{(9)} & --- & ---\\
--- & --- & 12.9\tnote{(10)} & 61.6\tnote{(10)}\\
--- & --- & 12.3\tnote{(11)} & 78.0\tnote{(11)}\\
\hline
\end{tabular}
\begin{tablenotes}
\item[1] PISO DM halo mass (M$_{\odot}$).
\item[2] PISO DM halo scale radius (kpc).
\item[3] NFW DM halo concentration at density contrast 200.
\item[4] NFW DM halo circular velocity at density contrast 200.
\item[5] \citet{Sicotte1997}.
\item[6] \citet{vandenBosch2001} with M/L=1.
\item[7] \citet{Swaters2003}.
\item[8] \citet{Spano2008}.
\item[9] This work. BKT fit to the H$\alpha$ and HI DM RCs.
\item[10] This work. NFW fit to the HI DM RC.
\item[11] This work. NFW fit to the H$\alpha$ DM RC.
\end{tablenotes}
\end{threeparttable}
\end{table} 

\section{Discussion}\label{sec:s12}

In this section we indicate possible limitations and issues related to the methodology presented in this paper that could in principle compromise part of the validity of some of the results of our analysis. The procedure adopted is based on some fundamental assumptions whose description we recapitulate next and whose validiy is discussed below. The foremost premises on which our approach is founded are the following: 

\begin{enumerate}

\item The circular motions dominate over non circular ones.\\ 

\item The stellar and gas tracers represent the total baryonic content of the analysed galaxies.\\ 

\item The DM constitutes the most plausible option to account for the observed discrepancy between the measured RC and the baryonic mass content.\\ 

\item The DM is a cold fluid that exchanges energy, linear and angular momenta with other mass components through collisionless gravitational interactions.\\ 

\end{enumerate}

The first hypothesis can be validated directly from the analysis of the observed velocity fields of UGC 8490 and UGC 9753 performed in section~(\ref{sec:s8}), where we have determined the best observed RC for both galaxies. The second hypothesis can be confirmed by the information relative to the detections of gas phases distinct from HI that can represent a significative and additional amount of mass for UGC 8490 and UGC 7953. We have partially addressed the latter question in section~(\ref{sec:s7}), reporting the results of the available studies that detected the presence of molecular hydrogen in the disc of both galaxies. 

The first and the second hypotheses seem to be interconnected, because the study of non circular motions in a galaxy rely, on one side, on the possibility to observe some gaseous or stellar component suitable to trace the galactic gravitational potential, and on the other side, on the possibility to properly separate the circular motions from significant velocity dispersions. In the particular case of UGC 8490 and UGC 9753, we perform the HI velocity field analysis in section~(\ref{sec:s8}), creating a synthetic circular velocity field through the GIPSY task VELFI and subtracting the model circular velocity field from the observed one to obtain an estimation of the magnitude of possible non circular motions. As already mentioned, the non circular motions in the HI disc of UGC 8490 and UGC 9753 are of the order of $\lesssim$ 20 km s$^{-1}$ or less, and as a consequence do not represent a great disturbance to the overall rotation of both galaxies (see figure~(\ref{fig4})). The H$\alpha$ velocity fields of both galaxies have been carefully analysed by \citet{Epinat2008}. The authors do not report severe departure from circular motions and the amplitude of the H$\alpha$ RCs is comparable with the corresponding circular velocity peaks of the HI RCs. From the observed H$\alpha$ and HI velocity fields of UGC 8490 and UGC 9753, the general conclusion is that circular rotation predominates over non circular motions. 

The second hypothesis is based first on the SPS analysis performed for the stellar discs of UGC 8490 and UGC 9753 and then on the detection of a sufficient quantity of molecular hydrogen that can contribute significantly to the total baryonic mass of both galaxies. The procedure adopted to accomplish the SPS analysis of UGC 8490 and UGC 9753 (see RP17) takes into account all the stellar mass and remnants (white dwarfs, neutron stars, or black holes) whose mass could constitute a substantial portion of the total baryonic mass of the analysed galaxies, and that otherwise would elude observational detection because of their very low luminosity. UGC 8490 does not host molecular hydrogen (see section~(\ref{sec:s7})) and no observational analysis seems to exist in the current literature that reports the detection of molecular hydrogen for UGC 9753. In general the search for cold molecular hydrogen in galaxies is a complicated task, principally due to the strong dependence of the conversion factor between the CO intensity and the H$_2$ mass on metallicity \citep{Arnault1988, Arimoto1996}. From the assumptions used to perform the SPS analysis of UGC 8490 and UGC 9753 and the observational evidences available in the literature to account for the molecular gas content, we conclude that the total baryonic mass of UGC 8490 and UGC 9753 derived in the present study represents a good approximation to the actual stellar and gas content of the analysed galaxies.

\subsection{Alternatives to the DM model}\label{sec:s13}

We have considered throughout this work that DM represents a plausible explication to the discrepancy among the observed RCs of UGC 8490 and UGC 9753 and the total baryonic mass of those two galaxies. In the available literature some authors fit the observed RC of UGC 8490 with different gravity models. As an alternative to the cold DM paradigm, in the following, we describe the main findings of those studies to contrast their diverse implications with the significance of the DM fit results. \citet{Swaters2010} (SW10) analysed if the predictions of the Modified Newtonian Dynamics (MOND), first introduced by \citet{Milgrom1983}, could reproduce the total gravitating mass of UGC 8490, adjusting the combined H$\alpha$ and HI RCs of this galaxy. The result was that MOND does not fit properly the full circular velocity magnitude of the studied RC, even though the authors comment that the RC form and circular velocity extent could not be so much reliable, due to the presence of a significant deformation of the HI disc. \citet{OBrien2012} used the weak field limit of the conformal gravity theory in the form proposed by \citet{Mannheim2006} to fit the RCs of 27 galaxies among which UGC 8490 is included, employing the same RC of SW10. The outcome of the quoted study was that the conformal gravity theory accounts in a satisfactory manner for the total gravitating mass of UGC 8490, although to obtain an adequate fit to the RC the authors increased the estimated inclination of this galaxy by 10$\degr$. \citet{Rodrigues2014} compared three different models, namely the NFW, MOND and the theory of renormalization group effects on general relativity (RGP) whose action and field equations were devised by \citet{Reuter2004}. The results of the RC fit of UGC 8490 favoured the NFW DM halo, nevertheless the NFW fit has one more free parameter with respect to RGP and two more free parameters if confronted with MOND. In the current literature we do no find any research work that examine the possibility of modified gravity or other gravity theories to explain the excess of mass of UGC 9753 with respect to its total baryonic mass. In summary the usage of different theoretical frameworks to explain the dissimilarity among the total and the baryonic mass in galaxies is a necessary tool to enlarge the theoretical bases of analysis and interpretation of that problem, despite of the difficulties to compare several different results. The effort is important and could generate new insights and perspectives on that topic.           
 
\section{Conclusions}\label{sec:s14}

In this article we present a general expression for the circular velocity of a test particle in a Newtonian 3D gravitational potential capable to be applied to any particular system of curvilinear coordinates that admits orthogonalization. It is important to emphasise that a proper application of the circular velocity formula given in this analysis requires that the matrix of the metric tensor coefficients can be diagonalised, however the latter does not represent an actual restriction to the applicability of the established circular velocity relation, because it relies on successive coordinates transformations and basic rules of matrix algebra. In addition any transformation applied to the original curvilinear coordinates system does not affects the actual mass distribution of the galaxy under study, it simply redefines the observer position with respect to the entire set of indirect observables (e.g. gravitational potential and spatial density) that enter Poisson equation and allow the determination of the presented circular velocity formula.  

We employ stellar population synthesis models and the HI total density map to obtain the stellar and gas (HI+He+metals) RCs of UGC 8490 and UGC 9753. The DM RCs of both galaxies originated by the subtraction of the stellar and the stellar plus gas RCs from the observed H$\alpha$ and HI RCs of both galaxies, respectively. As an application we employ the novel circular velocity relation to fit the DM RCs of UGC 8490 and UGC 9753, in order to obtain the DM and total mass of both galactic systems, considering a spherical, prolate and oblate spheroidal mass distributions for the DM halos of both galaxies. The most important outcomes of the present work can be summarized in the three points below:

\begin{itemize}
\item Cored EIN and BKT DM halos produce better results than the cored STD and cuspy NFW and DCN DM halos, regardless of the geometry of the density distribution.
\vspace{0.4cm}
\item In general, spherical and oblate cored DM halos perform better than the corresponding prolate cored DM models. 
\vspace{0.4cm}
\item The DM halos of UGC 8490 and UGC 9753 are better represented by cored oblate DM models than the correspondent cored prolate DM halos.
\end{itemize}

The latter result is in total agreement with the findings of recent cosmological numerical simulations about the dynamical evolution of low mass (M$_{\mathrm{DM}}$ $\lesssim$ 10$^{12}$ M$_{\odot}$) triaxial DM halos at the present time \citep{Vega2017}. In general the spherical, prolate and oblate fits to the H$\alpha$ and HI RCs of UGC 8490 and UGC 9753 generate DM and total masses consistent with the corresponding DM halos scale radius, the particular choice of the prolate and oblate spheroidal symmetry axes, the related oblateness and prolateness parameters and the baryonic mass distribution. In addition the findings of the present work, for a spherical mass distribution, are consistent with the results of analogous studies.  

\section*{Acknowledgements}

We acknowledge the anonymous referee for the constructive report that improves the presentation of the article. M. R. acknowledges financial support from DGAPA-PAPIIT (UNAM) IN103116 and CONACYT CY-253085 research projects. This research has made use of the NASA/IPAC Extragalactic Database (NED) which is operated by the Jet Propulsion Laboratory, California Institute of Technology, under contract with the National Aeronautics and Space Administration. We acknowledge the usage of the HyperLeda database (http://leda.univ-lyon1.fr). This research has made use of the SIMBAD database, operated at CDS, Strasbourg, France. Based on observations made with the NASA/ESA Hubble Space Telescope, and obtained from the Hubble Legacy Archive, which is a collaboration between the Space Telescope Science Institute (STScI/NASA), the Space Telescope European Coordinating Facility (ST-ECF/ESA) and the Canadian Astronomy Data Centre (CADC/NRC/CSA). Funding for the SDSS and SDSS-II has been provided by the Alfred P. Sloan Foundation, the Participating Institutions, the National Science Foundation, the U.S. Department of Energy, the National Aeronautics and Space Administration, the Japanese Monbukagakusho, the Max Planck Society, and the Higher Education Funding Council for England. The SDSS Web Site is http://www.sdss.org/.
The SDSS is managed by the Astrophysical Research Consortium for the Participating Institutions. The Participating Institutions are the American Museum of Natural History, Astrophysical Institute Potsdam, University of Basel, University of Cambridge, Case Western Reserve University, University of Chicago, Drexel University, Fermilab, the Institute for Advanced Study, the Japan Participation Group, Johns Hopkins University, the Joint Institute for Nuclear Astrophysics, the Kavli Institute for Particle Astrophysics and Cosmology, the Korean Scientist Group, the Chinese Academy of Sciences (LAMOST), Los Alamos National Laboratory, the Max-Planck-Institute for Astronomy (MPIA), the Max-Planck-Institute for Astrophysics (MPA), New Mexico State University, Ohio State University, University of Pittsburgh, University of Portsmouth, Princeton University, the United States Naval Observatory, and the University of Washington. We have made use of the WSRT on the Web Archive. The Westerbork Synthesis Radio Telescope is operated by the Netherlands Institute for Radio Astronomy ASTRON, with support of NWO. The WHISP observation were carried out with the Westerbork Synthesis Radio Telescope, which is operated by the Netherlands Foundation for Research in Astronomy (ASTRON) with financial support from the Netherlands Foundation for Scientific Research (NWO). The WHISP project was carried out at the Kapteyn Astronomical Institute by J. Kamphuis, D. Sijbring and Y. Tang under the supervision of T.S. van Albada, J.M. van der Hulst and R. Sancisi. This research has made use of the VizieR catalogue access tool, CDS, Strasbourg, France. The original description of the VizieR service was published in A$\&$AS 143, 23.







\appendix

\section{General circular velocity relation from Poisson equation}\label{sec:Aa1}

A reasonable and elementary interpretation of the 3D Poisson equation, at an specific time, is that a particular spatial density distribution originates a corresponding gravitational field that, through the interactions among its constituent mass elements, defines the concavity (minima) and convexity (maxima) of the gravitational potential boundary. In the following we consider a general system of curvilinear coordinates $u_i = (u_1,u_2,u_3)$ and we ideally subdivide the total 3D spatial density of the massive object under study into volume density subregions capable to produce their proper self-gravity. The volume density $\xi_{ij}$ and gravitational potential $\Psi_{ij}$ can be expressed as 3$\times$3 matrices, with the following elements:

\begin{align}\label{eqn:ap1}
\xi_{11} &=\rho(u_1)+\rho(u_2,u_3) & \xi_{12} &= \rho(u_1,u_2)+\rho(u_3)\nonumber\\
\xi_{13} &= \rho(u_1,u_3)+\rho(u_2) & \xi_{21} &= \rho(u_2,u_1)+\rho(u_3)\nonumber\\
\xi_{22} &= \rho(u_2)+\rho(u_3,u_1) & \xi_{23} &= \rho(u_2,u_3)+\rho(u_1)\nonumber\\
\xi_{31} &= \rho(u_3,u_1)+\rho(u_2) & \xi_{32} &= \rho(u_3,u_2)+\rho(u_1)\nonumber\\
\xi_{33} &= \rho(u_3)+\rho(u_2,u_1) & \Psi_{11} &= \Phi(u_1)+\Phi(u_2,u_3)\nonumber\\
\Psi_{12} &= \Phi(u_1,u_2)+\Phi(u_3) & \Psi_{13} &= \Phi(u_1,u_3)+\Phi(u_2)\nonumber\\
\Psi_{21} &= \Phi(u_2,u_1)+\Phi(u_3) & \Psi_{22} &= \Phi(u_2)+\Phi(u_3,u_1)\nonumber\\
\Psi_{23} &= \Phi(u_2,u_3)+\Phi(u_1) & \Psi_{31} &= \Phi(u_3,u_1)+\Phi(u_2)\nonumber\\
\Psi_{32} &= \Phi(u_3,u_2)+\Phi(u_1) & \Psi_{33} &= \Phi(u_3)+\Phi(u_2,u_1)
\end{align}

It is worth to notice that the results obtained in this section are independent of the particular choice of the entries of the volumetric density and gravitational potential arrays. The correspondent 3D Poisson equation, following \citet{Arfken2005}, can be expressed according to the relation:

\begin{align}\label{eqn:ap2}
\nabla^2\Phi(u_1, u_2, u_3) &= \frac{1}{\sqrt{g}} \left[\sum_{i,j=1}^3 \frac{\partial}{\partial u_i}\left(\sqrt{g}g^{ij}\frac{\partial \Psi_{ij}}{\partial u_j}\right)\right]\nonumber\\ 
&= 4\pi G \sum_{i,j=1}^3\xi_{ij} = 4\pi G \rho(u_1, u_2, u_3)
\end{align}

\noindent where $g^{ij}$ is the inverse of the Euclidean metric tensor matrix, and $g$ is a scalar function of the tensor matrix coefficients. We consider that the selected system of curvilinear coordinates is orthogonal, i.e. the tensor matrix is diagonal, leaving unaltered the matrices $\Psi_{ij}$ and $\xi_{ij}$. Let us notice that we represent the gravitational potential and volume density as generic matrices, without special properties such as symmetry, antisymmetry or orthogonality, this fact enforces the generality of our treatment. The resultant Poisson equation contains only diagonal terms of the gravitational potential, as can be demonstrated developing the sum on the left-hand side of equation~(\ref{eqn:ap2}) with $g^{ij}$ replaced by $g^{ii}$. On the right-hand side of equation~(\ref{eqn:ap2}) each non diagonal element of the volume density matrix can be dropped, because the corresponding gravitational potential second derivatives disappear, and as a consequence, the off-diagonal terms of the spatial density, that correspond to the vanishing second derivatives of the gravitational potential, can be omitted. In addition the first derivatives $\frac{\partial \Phi(u_2,u_3)}{\partial u_1}$, $\frac{\partial \Phi(u_1,u_3)}{\partial u_2}$, $\frac{\partial \Phi(u_2,u_1)}{\partial u_3}$ and the sum $4\pi G\left[\rho(u_2,u_3)+\rho(u_1,u_3)+\rho(u_2,u_1)\right]$ goes to zero and the resulting Poisson equation reads:

\begin{align}\label{eqn:ap3}
\nabla^2\Phi(u_1, u_2, u_3) &= \frac{1}{\sqrt{g}} \left[\frac{\partial}{\partial u_1}\left(\sqrt{g}g^{11}\frac{\partial \Phi(u_1)}{\partial u_1}\right)\right]+\nonumber\\
& +\frac{1}{\sqrt{g}} \left[\frac{\partial}{\partial u_2}\left(\sqrt{g}g^{22}\frac{\partial \Phi(u_2)}{\partial u_2}\right)\right]+\nonumber\\
& +\frac{1}{\sqrt{g}} \left[\frac{\partial}{\partial u_3}\left(\sqrt{g}g^{33}\frac{\partial \Phi(u_3)}{\partial u_3}\right)\right]\nonumber\\
&= 4\pi G \left[\rho(u_1)+\rho(u_2)+\rho(u_3)\right]\nonumber\\ 
&= 4\pi G \rho(u_1, u_2, u_3)
\end{align}    

Let us clarify that the volume density and gravitational potential matrices remain identical to those originally defined, however the observer can solely measure those terms that appear in equation~(\ref{eqn:ap3}), due to the transformation of the metric tensor matrix that is equivalent to an alteration of the observer reference frame (i.e. $g_{ij}=\frac{\partial x_k}{\partial u_i}\frac{\partial x_k}{\partial u_j}=0$), although, the physical quantities of the system under analysis remain unchanged. We can divide equation~(\ref{eqn:ap3}) into three parts on both the left and right-hand sides. The split of equation~(\ref{eqn:ap3}) is justified observationally because, to first approximation, in a real galaxy it is possible to separate the circular from the non circular motions, in particular we are interested in the first term that produces the observed rotation law, specifically:

\begin{align}\label{eqn:ap4}
\nabla^2\Phi(u_1) &= \frac{1}{\sqrt{g}} \left[\frac{\partial}{\partial u_1}\left(\sqrt{g}g^{11}\frac{v^2_c(u_1)}{u_1}\right)\right] = 4\pi G \rho(u_1)
\end{align} 

\noindent where we have considered the expression of the circular velocity of a test particle in a central gravitational potential $\frac{\partial\Phi(u_1)}{\partial u_1} = \frac{v^2_c(u_1)}{u_1}$ \citep{Binney2008}, that allows us to obtain the desired circular velocity relation:

\begin{align}\label{eqn:ap5}
v^2_c(U_1) &= \frac{4\pi G U_1}{\sqrt{g}g^{11}} \int_0^{U_1} \rho(u_1)\sqrt{g}du_1 
\end{align}

\noindent where the quantities $u_1$ and $U_1$ represent radial coordinates. It is important to emphasise that we do not attempt to solve the Poisson equation, and that it is not even necessary to achieve our purpose, instead we provide an integral (or a differential) formula based on the Poisson equation with the solely assumption that the metric tensor matrix can be reduced to orthogonal form. As we have already explained in section~(\ref{sec:s3}) we are interested in a tensorial form of the circular velocity relation, because of the broader range of applicabiliy of equation~(\ref{eqn:ap5}) to several astrophysical contexts involving fluid element kinematics in very different geometrical configurations. In section~(\ref{sec:s3}) we highlight some of the possible applications of this relation and also the principal limitations. In section~(\ref{sec:s4}) we apply the circular velocity relation to the specific prolate and oblate spheroidal geometries, and in Appendix~(\ref{sec:Bb1}) we detail the more relevant steps of the prolate and oblate application of equation~(\ref{eqn:ap5}). 

\section{Prolate and oblate spheroidal geometries}\label{sec:Bb1}

We consider the prolate $[x_i]_p$ and oblate $[x_i]_o$ spheroidal curvilinear coordinates transformations expressed by the relations:

\begin{align}\label{eqn:bp1}
[x_1]_p &= ra\cos{\psi}\sin{\eta} & [x_1]_o &= rb\cos{\psi}\sin{\eta}\nonumber\\
[x_2]_p &= ra\sin{\psi}\sin{\eta} & [x_2]_o &= rb\sin{\psi}\sin{\eta}\nonumber\\
[x_3]_p &= rb\cos{\eta}           & [x_3]_o &= ra\cos{\eta}
\end{align}

We use the Lame coefficients formalism (e.g. \citet{Aramanovich1961}, chapter IV, pages 99-134) to relate the metric tensor matrix elements to the partial derivative of the prolate and oblate spheroidal curvilinear coordinates of equation~(\ref{eqn:bp1}). In general the connection between the Lame symbols, the metric tensor coefficients and the adopted coordinates transformation is provided by the expressions:

\begin{align}\label{eqn:bp2}
l^2_i &=g_{ii}=\sum_{j=1}^3 \left(\frac{\partial x_j}{\partial u_i}\right)^2 & l_{lk} &=g_{lk}=\sum_{j=1}^3 \frac{\partial x_j}{\partial u_l}\frac{\partial x_j}{\partial u_k},
\end{align}

\noindent where for our particular prolate and oblate spheroidal coordinates we have $x_j=(x_1, x_2, x_3)$ and $u_i=(r,\psi,\eta)$, and the metric tensor coefficients are the following:

\begin{align}\label{eqn:bp3} 
[g_{11}]_p &= a^2\sin^2{\eta}+b^2\cos^2{\eta} & [g_{12}]_p &=0\nonumber\\ 
[g_{13}]_p &= r(a^2-b^2)\cos{\eta}\sin{\eta} & [g_{21}]_p &=0\nonumber\\
[g_{22}]_p &= r^2a^2\sin^2{\eta} & [g_{23}]_p &=0\nonumber\\
[g_{31}]_p &= r(a^2-b^2)\cos{\eta}\sin{\eta} & [g_{32}]_p &=0\nonumber\\
[g_{33}]_p &= r^2\left[a^2\cos^2{\eta}+b^2\sin^2{\eta}\right]\nonumber\\
[g_{11}]_o &= b^2\sin^2{\eta}+a^2\cos^2{\eta} & [g_{12}]_o &=0\nonumber\\ 
[g_{13}]_o &= r(b^2-a^2)\cos{\eta}\sin{\eta} & [g_{21}]_o &=0\nonumber\\
[g_{22}]_o &= r^2b^2\sin^2{\eta} & [g_{23}]_o &=0\nonumber\\
[g_{31}]_o &= r(b^2-a^2)\cos{\eta}\sin{\eta} & [g_{32}]_o &=0\nonumber\\
[g_{33}]_o &= r^2\left[b^2\cos^2{\eta}+a^2\sin^2{\eta}\right].
\end{align}

From equations~(\ref{eqn:bp3}) it can be easily demonstrated that the metric tensor matrix of the prolate and oblate spheroids is a symmetric non singular matrix. The purpose of this appendix is to derive equations~(\ref{eqn:ee2}) and~(\ref{eqn:ee3}) from equation~(\ref{eqn:ee1}), and in particular obtaining the quantities $g$ and $g^{11}$. In appendix~(\ref{sec:Cc1}) we perform the diagonalization of the metric tensor matrix and we obtain $\sqrt{g}=\left[g_{22}\left(g_{11}g_{33}-g^2_{13}\right)\right]^{\frac{1}{2}}$. For the prolate spheroid we have $\sqrt{g}=r^2 a^2 b \sin{\eta}$, whereas for the oblate spheroid we obtain $\sqrt{g}=r^2 b^2 a \sin{\eta}$. We determine the quantity $g^{11}$ calculating the inverse of the metric tensor matrix defined by equation~(\ref{eqn:bp3}) (see appendix~(\ref{sec:Dd1}) for details). From the computation of the inverse of the metric tensor matrix we obtain $g^{11}=\frac{g_{33}}{\left[g_{11}g_{33}-g^2_{13}\right]}$, in particular for the prolate spheroid we have $g^{11}=\frac{\cos^2{\eta}}{b^2}+\frac{\sin^2{\eta}}{a^2}$, whereas for the oblate spheroid we obtain $g^{11}=\frac{\cos^2{\eta}}{a^2}+\frac{\sin^2{\eta}}{b^2}$. We replace in equation~(\ref{eqn:ee1}) the quantities $\sqrt{g}$ and $g^{11}$ defined above for the prolate and oblate spheroid separately and we obtain the corresponding circular velocity expressions reported in section~(\ref{sec:s4}). The corresponding spherical relation can be obtained by applying the same methodology and replacing $a=b=1$ in equation~(\ref{eqn:bp1}).  

\section{Diagonalization of the symmetric metric tensor matrix}\label{sec:Cc1}

In this addendum we diagonalize the symmetric metric tensor matrix, considering the widely known fact of basic geometry of vectorial spaces, that states that the characteristic polynomial of a squared symmetric matrix is equivalent to the characteristic polynomial of every
endomorphism applied to the original matrix (e.g. \citet{Gantmacher1959}, chapter IV, pages 76-94). The resultant diagonal matrix is an endomorphism of the original symmetric matrix and as a consequence the characteristic polynomial of the original matrix is equivalent to the characteristic polynomial of the diagonal matrix. The symmetric metric tensor matrix $G_{ij}$ and the corresponding diagonalised matrix $G_{ii}$ are defined in the following manner:   

\begin{align}\label{eqn:cp1}
G_{ij} &=
\begin{pmatrix}
g_{11} & 0 & g_{13}\\   
0    & g_{22} & 0\\
g_{31} & 0 & g_{33}\\   
\end{pmatrix} & G_{ii} &=
\begin{pmatrix}
G_{11} & 0 & 0\\
0 & G_{22} & 0\\
0 & 0 & G_{33}\\
\end{pmatrix}
\end{align}

\noindent where $g_{13}=g_{31}$ according to equation~(\ref{eqn:bp3}) and $G_{11}$, $G_{22}$, $G_{33}$ represent unknown quantities. Computing the characteristic polynomial of the symmetric $p_c(u)$ and the diagonal matrix $p^{\prime}_c(u)$ expressed by equation~(\ref{eqn:cp1}) we have the following relations:

\begin{align}\label{eqn:cp2}
p_c(u) &= (u-g_{22})\left[u^2-(g_{11}+g_{33})u+g_{11}g_{33}-g^2_{13}\right]\nonumber\\
p^{\prime}_c(u) &= (u-G_{22})\left[u^2-(G_{11}+G_{33})u+G_{11}G_{33}\right].\nonumber\\
\end{align}

Considering the condition $p_c(u)=p^{\prime}_c(u)$ we have immediately that $G_{22}=g_{22}$ and:

\begin{align}\label{eqn:cp3}
g_{11}+g_{33} &= G_{11}+G_{33}\nonumber\\ 
g_{11}g_{33}-g^2_{13} &= G_{11}G_{33}.
\end{align}

From equation~(\ref{eqn:cp3}) we obtain the quadratic equation $G^2_{33}-G_{33}(g_{11}+g_{33})+g_{11}g_{33}-g^2_{13}=0$ that allows us to find $G_{33}$ and subsequently we employ again equation~(\ref{eqn:cp3}) to determine $G_{11}$. The solutions $G_{33}=\frac{1}{2}\left[(g_{11}+g_{33})\pm\sqrt{(g_{11}-g_{33})^2+4g^2_{13}}\right]$ and $G_{11}=\frac{1}{2}\left[(g_{11}+g_{33})\mp\sqrt{(g_{11}-g_{33})^2+4g^2_{13}}\right]$ resolve the problem of diagonalization of the original symmetric matrix $G_{ij}$ and it is straightforward to verify that these four roots satisfy the constraints expressed in equation~(\ref{eqn:cp3}) and consequently $g=G_{11}G_{22}G_{33}=g_{22}\left[g_{11}g_{33}-g^2_{13}\right]$ .  

\section{Inverse matrix calculation}\label{sec:Dd1}

The computation of the inverse matrix of the metric tensor matrix $G_{ij}$, expressed by equation~(\ref{eqn:cp1}), is important to establish the interrelations among the elements of the matrices $G_{ij}$ and $G^{ij}$. The coefficient $g^{11}$, for instance, enters in the definition of equation~(\ref{eqn:ee1}) and as a consequence is relevant for our analysis. From the definition $G_{ij}G^{ij}=I_3$, where $I_3$ is the $3\times3$ identity matrix, we derive the following elements for the matrix $G^{ij}$:

\begin{align}\label{eqn:dp1}
g^{11} &= \frac{g_{33}}{g_{11}g_{33}-g^2_{13}} & g^{12} &=0\nonumber\\
g^{13} &= \frac{g_{13}}{g^2_{13}-g_{11}g_{33}} & g^{21} &=0\nonumber\\
g^{22} &= \frac{1}{g_{22}} & g^{23} &= 0\nonumber\\
g^{31} &= \frac{g_{31}}{g^2_{31}-g_{11}g_{33}} & g^{32} &=0\nonumber\\
g^{33} &= \frac{g_{11}}{g_{11}g_{33}-g^2_{13}}
\end{align}

\noindent where $g^{13}=g^{31}$ (because $g_{13}=g_{31}$), and consequently $G^{ij}$ represents a symmetric non singular matrix as the original matrix $G_{ij}$. It is important to note that (see equation~(\ref{eqn:dp1})) the relations between the elements of $G^{ij}$ and
$G_{ij}$ are not simply expressed as the inverse of the matrix coefficients, except for the matrix element $g_{22}$, but are more elaborated expressions that involve the off-diagonal terms $g_{13}$ and $g_{31}$, due to the fact that the original matrix $G_{ij}$ is not diagonal. As we have already mentioned, the inverse matrix calculation is a necessary step to particularize the general circular velocity relation expressed by equation~(\ref{eqn:ee1}).  

\section{Axes of symmetry of prolate and oblate spheroids: constraining relations}\label{sec:Ee1}

In this appendix we determine the relations of the major and minor axes of symmetry of the prolate and oblate spheroidal mass distributions considered to describe the DM halos of UGC 8490 and UGC 9753. We assume that DM is a collisionless fluid whose total energy is conserved. In the following we detail the derivation of the expressions of the minor and major axes of symmetry for the prolate spheroidal geometry, the corresponding oblate spheroidal quantities are obtained through the application of identical arguments. We first consider an hypothetical DM particle of mass $m$ moving on a prolate spheroidal orbit on the surface of the corresponding prolate spheroid. At the present time $t_0$ the circular velocity of such a particle can be expressed as $\left[v_c(r_p)\right]_p=\frac{n_1\pi r_p}{t_0}$, where $0 < n_1 \leq 2$ and $n_1\pi r_p$ represents the prolate spheroidal arc drawn by the prolate circular motion of the considered DM particle. At the time $t_0$
we are in a matter dominated universe where $t_0=\frac{2}{3H_0}$ (e.g. \citet{Coles2002}) and $H_0$ is the current Hubble constant expressed in units of s$^{-1}$. The circular velocity as a function of $H_0$ is $\left[v_c(r_p)\right]_p=\frac{3 n_1\pi r_p H_0}{2}$ and the corresponding kinetic energy is provided by the relation $\left[E_{kin}\right]_p=\frac{9n^2_1\pi^2 r^2_p H^2_0 m}{8}$. The gravitational potential energy of the studied DM particle is expressed as $\left[E_{pot}\right]_p=-\frac{GMm}{r_p}$, where $M$ is the total mass of the prolate spheroidal geometry under analysis and $G$ is the Newtonian constant of gravitation in units of M$^{-1}_{\odot}$ kpc km$^2$ s$^{-2}$ . The condition of energy conservation implies $\left[E_{kin}\right]_p+\left[E_{pot}\right]_p=0$, and setting $r_p=\frac{r}{a}$ where $r$ represent the RC radius and $a$ the minor axis of symmetry of the prolate spheroidal mass distribution, we have for the minor and major axes of symmetry:

\begin{align}\label{eqn:ep1}
a &= \left(\frac{9 n^2_1 \pi^2 H_0^2 r^3_i}{8 G M_i}\right)^{\frac{1}{3}}  & b &= \frac{3 M_i}{4 a^2 \pi \rho_i r^3_i}
\end{align} 

\noindent where $M_i$, $\rho_i$ and $r_i$ represent the prolate spheroidal initial mass, volume density and radius. In equation~(\ref{eqn:ep1}) $G$ is expressed in units of M$^{-1}_{\odot}$ kpc$^3$ s$^{-2}$. The expression for the major axis of symmetry of the prolate spheroid was obtained through the relation of mass given by equation~(\ref{eqn:ee4}) with a constant initial density, i.e., $M_i=\frac{4}{3}\pi a^2 b \rho_i r^3_i$. We derive equivalent relations for the minor and major axes of symmetry of the oblate spheroidal mass distribution applying the same reasoning. In the fitting process of the DM H$\alpha$ and HI RCs of UGC 8490 and UGC 9753 at each radial step the DM mass and radius are increased up to the convergence of the solution that produces the final quantities. The condition of energy conservation can be applied at every radial step beginning from the initial DM mass and radius up to the last DM mass and radius. This is equivalent to subdivide the prolate and oblate spheroids in concentric prolate and oblate spheroidal shells and computing for each shell the enclosed mass, having established the condition of energy conservation. The latter is an intrinsic property of the DM fluid, independent of the geometric configuration considered. The constant values of the minor and major axes of symmetry of the prolate and oblate spheroids utilized in the present study and reported in tables~(\ref{tab:tb4}) and~(\ref{tab:tb5}) are obtained by applying equation~(\ref{eqn:ep1}) with the appropriate values of mass, volume density and radius, and correspond to the actual values used in the fits of the H$\alpha$ and HI DM RCs.  


\bsp	
\label{lastpage}
\end{document}